%% file: alto.tex
\documentclass[10pt,journal,compsoc]{IEEEtran}

\input{text/commands}

\ifCLASSOPTIONcompsoc
  \usepackage[nocompress]{cite}
\else
  \usepackage{cite}
\fi

\begin{document}

\title{Accelerating Sparse Tensor Decomposition Using Adaptive Linearized Representation}

\author{
    Jan Laukemann, Ahmed E. Helal, S. Isaac Geronimo Anderson, Fabio Checconi, Yongseok Soh, Jesmin Jahan Tithi, Teresa Ranadive, Brian J Gravelle, Fabrizio Petrini, and Jee Choi
    
    \IEEEcompsocitemizethanks{
        \IEEEcompsocthanksitem Jan Laukemann is with Friedrich-Alexander-Universität Erlangen-Nürnberg and Erlangen National High Performance Computing Center. Email: jan.laukemann@fau.de.
        \IEEEcompsocthanksitem Ahmed E. Helal, Fabio Checconi, Jesmin Jahan Tithi, and Fabrizio Petrini are with Intel Labs. Email: \{ahmed.helal, fabio.checconi, jesmin.jahan.tithi, fabrizio.petrini\}@intel.com.
        \IEEEcompsocthanksitem Teresa Ranadive and Brian J Gravelle are with Laboratory for Physical Sciences. Email: \{tranadive, bjgrave\}@lps.umd.edu.
        \IEEEcompsocthanksitem S. Isaac Geronimo Anderson, Yongseok Soh, and Jee Choi are with University of Oregon. Email: \{igeroni3, ysoh, jeec\}@uoregon.edu.
    }
    \thanks{}
}
\markboth{}%
{Shell \MakeLowercase{\textit{et al.}}: Bare Demo of IEEEtran.cls for Computer Society Journals}

\IEEEtitleabstractindextext{%
\justify
\begin{abstract}
High-dimensional sparse data emerge in many critical application domains such as healthcare and cybersecurity.
To 
extract meaningful insights from massive volumes of these multi-dimensional data, scientists employ unsupervised analysis tools based on tensor decomposition (TD) methods. 
However, real-world sparse tensors exhibit highly irregular shapes and data distributions, 
which pose significant challenges for making efficient use of modern parallel processors. 
This study breaks the prevailing assumption that compressing sparse tensors into coarse-grained structures (i.e., tensor slices or blocks) or along a particular dimension/mode (i.e., mode-specific) is more efficient than keeping them in a fine-grained, mode-agnostic form.
Our novel sparse tensor representation, Adaptive Linearized Tensor Order (\ALTO), encodes tensors in a compact 
format that can be easily streamed from memory and is amenable to both caching and parallel execution. 
In contrast to existing compressed tensor formats, \ALTO constructs one tensor copy that is agnostic to both the mode orientation and the irregular distribution of nonzero elements.
To demonstrate the efficacy of \ALTO, we accelerate popular TD methods that compute the Canonical Polyadic Decomposition (CPD) model across different types of sparse tensors.
We propose a set of parallel TD algorithms that exploit the inherent data reuse of tensor computations to substantially reduce synchronization overhead, decrease memory footprint, and improve parallel performance.
Additionally, we 
characterize the major execution bottlenecks of 
TD methods on multiple generations of the latest Intel Xeon Scalable processors, including Sapphire Rapids CPUs, and introduce dynamic adaptation heuristics to automatically select the best algorithm based on the sparse tensor characteristics.
Across a diverse set of real-world 
data sets, 
\ALTO outperforms the state-of-the-art approaches, achieving more than an order-of-magnitude speedup over the best mode-agnostic formats.
Compared to the best mode-specific formats, which require multiple tensor copies, \ALTO achieves $5.1\times$ geometric mean speedup at a fraction 
($25\%$) of their storage costs. 
\revon Moreover, \ALTO obtains $8.4\times$ geometric mean speedup over the state-of-the-art memoization approach, which reduces computations by using extra memory, while requiring $14\%$ of its memory consumption.\revoff        
\end{abstract}

}

\maketitle

\IEEEdisplaynontitleabstractindextext

\IEEEpeerreviewmaketitle

\IEEEraisesectionheading{
	\section{Introduction}
	\label{sec:introduction}
}
\input{text/intro}

\input{text/background}

\input{text/approach}

\input{text/results}

\input{text/related}

\input{text/conclusion}
\ifCLASSOPTIONcaptionsoff
  \newpage
\fi

\bibliographystyle{IEEEtran}
\bibliography{IEEEabrv,alto}

\vspace{-2.5\baselineskip}
\begin{IEEEbiographynophoto}
{Jan Laukemann}
is a PhD student at the University of Erlangen-Nürnberg Erlangen and works for the Erlangen National High Performance Computing Center. Previously he worked as a Research Scientist at Intel Parallel Computing Lab. He focuses on application optimization and performance engineering for HPC systems and novel algorithms for scalable linear algebra, tensor decomposition and graph computations. His research interests primarily include x86 and non-x86 computer architectures, their performance behavior on the node level, and vectorization techniques. 
\end{IEEEbiographynophoto}

\vspace{-2.5\baselineskip}
\begin{IEEEbiographynophoto}{Ahmed E. Helal}
is a Research Scientist at the Intel Parallel Computing Lab in Santa Clara, CA, USA. He specializes in program analysis and transformation of irregular applications to optimize their performance on heterogeneous parallel architectures. His research interests include scalable algorithms for high-dimensional data analytics and sparse linear algebra.
Helal received his Ph.D. degree from the Bradley Department of Electrical and Computer Engineering at Virginia Tech.  
\end{IEEEbiographynophoto}

\vspace{-2.5\baselineskip}
\begin{IEEEbiographynophoto}{S. Isaac Geronimo Anderson}
S. Isaac Geronimo Anderson is a Ph.D. student advised by Jee Choi and Hank Childs in the department of Computer Science at University of Oregon. He has research interests in sparse tensors, performance portability, and HPC application interoperation. Isaac has interned at Sandia National Laboratories, Oak Ridge National Laboratory, and Cray (now Hewlett Packard Enterprise).
\end{IEEEbiographynophoto}

\vspace{-2.5\baselineskip}
\begin{IEEEbiographynophoto}{Fabio Checconi}
is a Research Scientist at the Intel Parallel Computing Lab in Santa Clara, CA, USA. His research interests include
large-scale graph algorithms and sparse linear algebra. Checconi received his Ph.D. degree in computer engineering from
Scuola Superiore S. Anna in Pisa, Italy.
\end{IEEEbiographynophoto}

\vspace{-2.5\baselineskip}
\begin{IEEEbiographynophoto}{Yongseok Soh}
is a Ph.D. student at the University of Oregon, where he is an active member of the High Performance Computing Research Group. 
His research primarily concentrates on enhancing the efficiency of sparse tensor factorization algorithms on HPC systems involving algorithmic and architectural adjustments and optimizations.
\end{IEEEbiographynophoto}

\vspace{-2.5\baselineskip}
\begin{IEEEbiographynophoto}{Dr. Jesmin Jahan Tithi}
is an AI Research Scientist and Tech Lead at Intel Corporation, specializing in algorithm engineering, high-performance computing (HPC), and hardware-software codesign. With a Ph.D. from Stony Brook University and a B.Sc. from Bangladesh University of Engineering and Technology, she has completed internships at Google, Intel, and the Pacific Northwest National Laboratory. Dr. Tithi has over 40 peer-reviewed publications and 12 published patents. 
\end{IEEEbiographynophoto}

\vspace{-2.5\baselineskip}
\begin{IEEEbiographynophoto}{Teresea Ranadive}
has been a researcher at the Laboratory for Physical Sciences since 2018.  She received a B.S. in Mathematics from St. Vincent College (2011), and an M.S. (2013) and Ph.D. (2016) in Applied Mathematics from the University of Maryland, Baltimore County.  Between 2016 and 2018, Dr. Ranadive was an Assistant Research Professor at Johns Hopkins University in the Applied Mathematics and Statistics department.  Her research interests include numerical optimization, tensor decomposition, and Ising spin glass models.  
\end{IEEEbiographynophoto}

\vspace{-2.5\baselineskip}
\begin{IEEEbiographynophoto}{Brian J. Gravelle}
Brian J Gravelle is a computer systems researcher at the Laboratory for Physical Sciences in Maryland, USA. He studied Computer Engineering at Gonzaga University before obtaining a PhD in Computer Science at the University of Oregon. In past research, he has evaluated the performance of various scientific and data analytics algorithms running on supercomputers and worked to improve the runtime of key kernels of computation. Currently, he studies new methods of improving the performance and programmability of distributed systems and GPUs with a particular focus on applications which rely on sparse linear algebra computation.
\end{IEEEbiographynophoto}

\vspace{-2.5\baselineskip}
\begin{IEEEbiographynophoto}{Fabrizio Petrini}
is a Senior Principal Engineer at the Intel Parallel Computing Lab in Santa Clara, CA, USA. His research interests include data-intensive algorithms for graph analytics and sparse linear algebra, Exascale computing, high-performance interconnection networks and novel architectures. He is the principal investigator of the TCStream CS project, Co-Principal investigator of the Intel PIUMA architecture, developed under the DARPA HIVE and SDH programs, and the upcoming Intel TIGRE system, as part of the IARPA AGILE program. Petrini received his Ph.D. degree in computer science from Pisa University, Italy. He is a Senior Member of IEEE.
\end{IEEEbiographynophoto}

\vspace{-2.5\baselineskip}
\begin{IEEEbiographynophoto}
{Jee Choi}
is an Assistant Professor in the Department of Computer Science at the University of Oregon. 
During his PhD, he worked on designing parallel and scalable algorithms for scientific applications, and modeling their performance and energy efficiency on the latest high-performance computing (HPC) systems. 
After graduation, he worked as a research staff member at the IBM T. J. Watson Research Center on designing and optimizing tensor decomposition algorithms for Big Data analytics.
\end{IEEEbiographynophoto}

\end{document}

%% file: text/commands.tex
\usepackage{amsmath}
\usepackage{amssymb}
\usepackage{color}
\usepackage{xcolor}
\usepackage{algorithm}
\usepackage{algpseudocode}

\let\oldReturn\Return
\renewcommand{\Return}{\State\oldReturn}
\usepackage{xspace}
\usepackage{bm}
\newcommand{\ALTO}{\mbox{{\sf ALTO}}\xspace}
\usepackage{listings}
\usepackage{graphicx}
\usepackage{pgfplots}
\usepgfplotslibrary{statistics}
\usepackage{pgfplotstable}
\pgfplotsset{compat=1.16}
\usepackage[font=small]{caption}
\usepackage{algorithm}
\usepackage{algpseudocode}
\usepackage{varwidth}
\usepackage{subcaption}
\usepackage{tikz}
\usetikzlibrary{patterns}
\usetikzlibrary{shapes}
\usepackage[nolist]{acronym}
\usepackage{paralist}
\usepackage{enumitem}
\usepackage{balance}
\usepackage{comment}
\usepackage{booktabs}
\usepackage{url}
\usepackage{ragged2e}
\usepackage{setspace}
\usepackage{hyperref}

\newcommand{\MATRIX}[1]{$\textbf{#1}$}

\newcommand{\VECELEM}[2]{${#1}_{#2}$}
\newcommand{\TENSOR}[1]{$\mathcal{#1}$}
\newcommand{\TENSORI}[1]{\mathcal{#1}}

\newcommand{\REALTWO}[2]{$\mathbb{R}^{#1\times #2}$}
\newcommand{\REALTHREE}[3]{$\mathbb{R}^{#1\times #2 \times #3}$}

\newcommand{\IGNORE}[1]{}

\newcommand{\GHZ}{\mbox{GHz}}

\newcommand{\TB}{\mbox{TB}}

\newcommand{\MiB}{\mbox{MiB}}

\newcommand{\Rlm}{Roof{}line model}

\begin{acronym}[MTTKRP]
\acro{CLX}{Cascade Lake-X}
\acro{ICX}{Ice Lake}
\acro{SPR}{Sapphire Rapids}
\acro{FROSTT}{the Formidable Repository of Open Sparse Tensors and Tools}
\acro{HaTen2}{Hadoop Tensor method for 2 decompositions}
\acro{ICPC}{Intel C++ Compiler}
\acro{MTTKRP}{matricized tensor times Khatri-Rao product}
\end{acronym}

\definecolor{greyback}{RGB}{248,248,248}
\definecolor{deepblue}{RGB}{43,131,186}
\definecolor{greenalt}{RGB}{35,139,69}
\definecolor{darkred}{RGB}{215,25,28}
\definecolor{darkorange}{RGB}{253,174,97}
\definecolor{light-gray}{gray}{0.85}
\definecolor{lighter-gray}{gray}{0.95}
\definecolor{applegreen}{rgb}{0.55, 0.71, 0.0}
\definecolor{asparagus}{rgb}{0.53, 0.66, 0.42}
\definecolor{babyblueeyes}{rgb}{0.63, 0.79, 0.95}
\definecolor{burntsienna}{rgb}{0.91, 0.45, 0.32}
\definecolor{deepcarmine}{rgb}{0.66, 0.13, 0.24}
\definecolor{lightcornflowerblue}{rgb}{0.6, 0.81, 0.93}
\definecolor{steelblue}{rgb}{0.27, 0.51, 0.71}
\definecolor{pastelblue}{rgb}{0.68, 0.78, 0.81}
\definecolor{pastelbrown}{rgb}{0.51, 0.41, 0.33}
\definecolor{pastelgray}{rgb}{0.81, 0.81, 0.77}
\definecolor{pastelgreen}{RGB}{184,216,190}
\definecolor{pastelmagenta}{rgb}{0.96, 0.6, 0.76}
\definecolor{pastelorange}{rgb}{1.0, 0.7, 0.28}
\definecolor{pastelpink}{rgb}{1.0, 0.82, 0.86}
\definecolor{pastelpurple}{rgb}{0.7, 0.62, 0.71}
\definecolor{pastelred}{rgb}{1.0, 0.41, 0.38}
\definecolor{pastelviolet}{rgb}{0.8, 0.6, 0.79}
\definecolor{pastelyellow}{rgb}{0.99, 0.99, 0.59}
\definecolor{darkblue}{rgb}{0.0, 0.0, 0.55}
\definecolor{darkgray}{rgb}{0.675, 0.655, 0.60}
\definecolor{light-gray}{gray}{0.85}
\definecolor{lighter-gray}{gray}{0.95}
\definecolor{lighter-gray2}{gray}{0.98}

\newcommand{\revon}{\color{black}}
\newcommand{\revoff}{\color{black}}

%% file: text/intro.tex
Tensors, the higher-order generalization of matrices, can naturally represent complex interrelations in multi-dimensional sparse data, which emerge in important application domains such as 
healthcare~\cite{ho2014marble, wang2015rubik}, cybersecurity~\cite{kobayashi2018extracting, fanaee2016tensor}, data mining~\cite{kolda2008scalable, papalexakis2016tensors}, and machine learning~\cite{anandkumar2014tensor, sidiropoulos2017tensor}. 
For example, one mode (or dimension) of a tensor may identify users while another mode details their demographic information, and their (potentially incomplete) ratings for a set of products~\cite{frostt}.
To effectively analyze such high-dimensional data, tensor decomposition (TD) is used to reveal their principal components, where each component represents a latent property.
One of the most popular TD models is the Canonical Polyadic Decomposition (CPD), which approximates a tensor as a sum of a finite number of rank-one tensors such that each rank-one tensor corresponds to a tensor component, or a latent property~\cite{KoBa09,Bader2007}. 
An important class of real-world, high-dimensional data sets is sparse tensors with \emph{non-negative count data}~\cite{10.1145/1921632.1921636,10.1145/1150402.1150445}, which encodes critical information, such as the number of packets exchanged across a network, or the number of criminal activities in a city. For these tensors, the CP Alternating Poisson Regression (CP-APR) algorithm is a powerful tool for detecting anomalies and group relations.
In contrast to other CPD algorithms that assume a Gaussian distribution for randomly distributed data (e.g., CP Alternating Least Squares, or CP-ALS), 
CP-APR~\cite{Chi12Tensors}
assumes a Poisson distribution which better describes the target 
count data. 

\input{fig/motivation_boxplot.tex}

Due to the curse of dimensionality, data become more dispersed as the number of modes increases. 
Hence, high-dimensional data sets typically suffer from highly irregular shapes and data distributions as well as unstructured and extreme sparsity, which make them challenging to represent efficiently. For instance,  Figure~\ref{fig:motivation} illustrates the spatial distribution of nonzero elements in a set of sparse tensors. It shows that the number of nonzero elements in a subspace, or a multi-dimensional block, can vary greatly (note the use of logarithmic scale). Furthermore, as the sparsity of tensors increases (e.g., \textsc{nell-1}, \textsc{amazon}, and \textsc{reddit} tensors), the likelihood of finding dense structures in the multi-dimensional space significantly decreases, leading to extremely small numbers of nonzero elements per block. Therefore, efficient execution of TD algorithms on modern parallel processors is challenging because of their low arithmetic intensity~\cite{choi2018blocking}, random memory access, workload imbalance~\cite{Kaya2015, blco_2022}, and data dependencies~\cite{choi2018blocking, blco_2022}. Moreover, TD algorithms 
require computations along \emph{every} mode, 
and realizing acceptable performance across all modes is difficult without using multiple mode-specific tensor copies.

Prior work on the CP-ALS algorithm improved the parallel performance of the matricized tensor times Khatri-Rao product (MTTKRP) kernel~\cite{choi2018blocking,Kaya2015,li2018hicoo,Smith2015,Liu2017}, which is the main performance bottleneck 
of the overall algorithm~\cite{Smith2015}.
The few performance studies conducted on the CP-APR algorithm focused primarily on performance portability~\cite{sparten} and streaming analysis~\cite{8547700}, rather than parallel performance optimization.
To the best of our knowledge, this paper presents the first in-depth analysis of the key performance bottlenecks of CP-APR, and significantly improves the parallel performance of both CP-ALS and CP-APR over prior state-of-the-art approaches by using a linearized mode-agnostic sparse tensor representation.

Additionally, the previous approaches relied on extending legacy sparse linear algebra formats and algorithms to tensor (multilinear) algebra problems~\cite{Bader2007, baskaran2012efficient, Smith2015, Smith2015a, Liu2017, phipps2019software, li2018hicoo, li2019efficient, nisa2019load, nisa2019efficient}. These techniques can be classified based on their compression of the nonzero elements into raw or compressed formats~\cite{alto_2021, alto_stream_2023}. Raw formats use simple list-based representations to keep the nonzero elements along with their multi-dimensional coordinates~\cite{chou2018format}. Hence, they are mode-agnostic and typically require one tensor copy to execute tensor operations along different modes. As a result, the list-based coordinate~(COO) format remains the de facto data structure for storing sparse tensors~\cite{chou2018format} in many 
tensor libraries (e.g., Tensor Toolbox~\cite{Bader2007}, Tensorflow~\cite{abadi2016tensorflow}, and Tensorlab~\cite{vervliet2016tensorlab}).   
However, due to their unprocessed nature, list-based formats suffer from significant parallel and synchronization overheads~\cite{Liu2017}.

Compressed tensor formats~\cite{Smith2015, smith2017accelerating, li2018hicoo, li2019efficient} use tree- or block-based structures to organize the nonzero elements, which may decrease the memory footprint of sparse tensors. However, since these approaches rely on finding clusters of nonzero elements in \textit{non-overlapping regions} of the multi-dimensional space to achieve compression, their efficacy depends on the spatial data distribution, which can be highly irregular and extremely sparse 
as demonstrated in Figure~\ref{fig:motivation}. 
Therefore, instead of reducing memory storage, compressed formats can introduce substantial memory overhead and degrade the parallel performance 
of TD algorithms
\cite{alto_2021}.

The most popular compressed format for TD algorithms is compressed sparse fiber~(CSF)~\cite{Smith2015}, which extends the classical compressed sparse row~(CSR) format to higher-order tensors using tree-like structures. However, CSF-based formats are mode-specific, where the arrangement of nonzero elements depends not only on the mode 
considered as the root of the index tree but also on the other modes at each subsequent tree level; therefore, they are typically efficient for \emph{only} that specific mode order. 
This leads to a trade-off between performance and memory, as storing multiple tensor copies, 
each arranged for a specific mode, yields the best performance at the cost of extra memory~\cite{smith2017accelerating}. For large-scale tensors, keeping multiple copies may be infeasible, especially for hardware accelerators with limited memory capacity (e.g., GPUs). 
Although current GPUs (e.g., H100) can have up to 80\,GB of device memory~\cite{choquette2023nvidia}, encoding large-scale tensors (e.g., \textsc{amazon}, \textsc{patents} and \textsc{reddit} tensors) using CSF requires hundreds of gigabytes of memory. 
In contrast, keeping only one tensor copy, 
arranged for an arbitrarily chosen mode order, yields the smallest memory footprint at the cost of sub-optimal performance~\cite{Smith2015a}. 
Alternatively, memoization schemes~\cite{li2017model, kurt2022sparsity} utilize CSF-based formats to reduce computations by keeping and reusing intermediate results across tensor modes, which require substantial storage that largely exceeds the memory consumption of tensor representations~\cite{kurt2022sparsity}.     
While block-based formats~\cite{li2018hicoo, li2019efficient} can be mode-agnostic, their storage requirements and parallel performance still depend on the spatial data distributions as well as the parameters of the blocking/tilling schemes~\cite{li2019efficient, alto_2021}, which are difficult to determine dynamically.      
 
To overcome these limitations, we present the \textit{Adaptive Linearized Tensor Order (\ALTO)} format. \ALTO is a mode-agnostic representation that maps a set of $N$-dimensional coordinates onto a single linearized index such that neighboring nonzero elements in the multi-dimensional space are close to each other in memory. This leads to a more cache friendly and memory-scalable tensor storage; that is, \ALTO utilizes the inherent data locality of sparse tensors and its storage scales with mode lengths, rather than the number of modes. Additionally, \ALTO enables a unified implementation of tensor algorithms that requires a single tensor copy to compute along all modes.

In contrast to prior compressed~\cite{Smith2015, li2018hicoo} and linearized~\cite{harrison2018high} TD approaches, we propose a set of parallel algorithms that leverage the \ALTO format to address the performance bottlenecks that have traditionally limited the scalability of sparse tensor computations. Our \ALTO-based algorithms generate perfectly balanced tensor partitions in terms of the number of nonzero elements; however, these partitions may divide the multi-dimensional space of a tensor into \emph{overlapping} regions (but each nonzero element belongs to a single partition). 
Thus, we devise data-aware adaptation heuristics that greatly 
improve parallel performance by locating the overlapping space between tensor partitions and 
selecting the best tensor traversal and conflict resolution method according to the inherent data locality of sparse tensors. Moreover, these heuristics choose between recomputing or reusing intermediate results, depending on the properties of the target tensor computations, to maximize the performance while reducing the overall memory footprint.
As a result, our \ALTO-based TD algorithms deliver substantial performance gains over prior state-of-the-art approaches, while allowing the processing of large-scale tensors. 
In summary, we make the following contributions:
\begin{itemize}[leftmargin=*]
\item We present \ALTO, a novel sparse tensor format for high-performance 
tensor algorithms. Unlike prior compressed 
formats, \ALTO uses a single (mode-agnostic) tensor representation that improves data locality, eliminates workload imbalance, and reduces memory usage, 
\textit{regardless of the data distribution
in the multi-dimensional space} ($\S$\ref{sec:approach}).
\item 
We propose efficient \ALTO-based parallel algorithms for the CP-ALS and CP-APR methods as well as input-aware adaptation heuristics to balance data reuse and
memory footprint while greatly reducing synchronization cost ($\S$\ref{sec:approach-algo}).
\item We conduct an in-depth performance analysis of the main TD kernels and compare against prior state-of-the-art across two generations of the latest Intel Xeon Scalable processors. The results
show that on an Intel Sapphire Rapids server, \ALTO-based TD algorithms achieve 
$25.3\times$ and $5.1\times$ 
geometric mean \emph{speedup} over the best mode-agnostic and mode-specific formats, respectively.
 Compared to the best memoization method that trades lower computation for higher memory usage, \ALTO attains $8.4\times$ geometric mean \emph{speedup}. Furthermore, \ALTO requires a fraction ($14\%$ to $25\%$) of the storage used by the state-of-the-art mode-specific and memoization approaches  ($\S$\ref{sec:perf}). 
\end{itemize}

%% file: fig/motivation_boxplot.tex
\pgfplotsset{   
    boxplot/.cd,
        every median/.style={draw=blue}
}

\begin{figure*}[htb]
\centering
\vspace*{-5pt}
\includegraphics{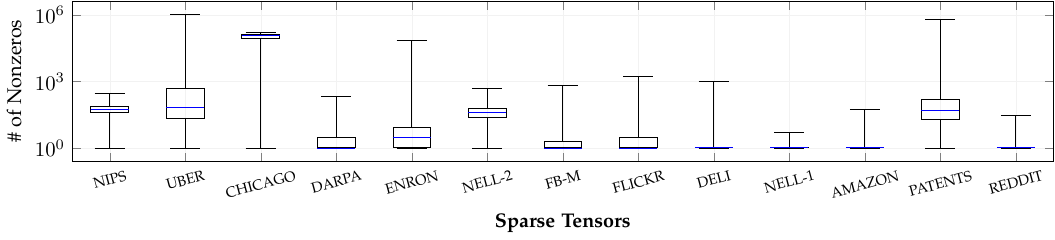}
	\vspace*{-10pt}
	\caption{A box plot of the data (nonzero elements) distribution across the multi-dimensional blocks (subspaces) of the hierarchical coordinate storage~\cite{li2018hicoo}.
	The multi-dimensional subspace size is 128$^{N}$, where $N$ is the number of dimensions (modes), as per prior work~\cite{sun2020sptfs}. The sparse tensors are sorted in an increasing order of their number of nonzero elements.
    \revon Sparsity is extremely high for many tensors (e.g., \textsc{nell-1}, \textsc{amazon}, and \textsc{reddit}) and vary greatly across tensors. \revoff}
	\vspace*{-8pt}
\label{fig:motivation}
\end{figure*}

%% file: text/background.tex
\section{Background}\label{sec:background}
This section summarizes popular tensor decomposition methods, sparse tensor formats, and related notations.
The survey by Kolda and Bader~\cite{KoBa09}  provides a more detailed discussion of tensor algorithms and their applications.

\vspace*{-7pt}
\subsection{Tensor Notations}
\vspace*{-2pt}
Tensors are $N$-dimensional arrays, where each element has an 
$N$-tuple index~$ \textbf{i}~=~(i_1, i_2, \ldots, i_N)$.
Each index coordinate $i_n$ locates a tensor element along the $n^{\mathrm{th}}$ dimension or mode, with $n \in \{1, 2, \ldots, N\}$ and $i_n \in \{1, 2, \ldots, I_n\}$.
Low-dimensional tensors include vectors, where $N = 1$, and matrices, where $N = 2$.
In general, a \emph{dense} $N$-dimensional (or a mode-$N$) tensor has $\prod_{n=1}^N I_n$ indexed elements.
A tensor is said to be sparse if the majority of its elements are zero.
The following notations are used in this paper:
\begin{enumerate}[leftmargin=*]
    \item Scalars are denoted by italicized lowercase letters (e.g. $a$).
    \item Vectors
    are denoted by bold lowercase letters (e.g. $\mathbf a$).
    \item Matrices are denoted by bold capital letters (e.g. $\mathbf A$).
    \item Higher-order tensors are denoted by Euler script letters (e.g. $\mathcal X$).
    \item Fibers are analogous to matrix rows and columns. 
    A mode-$n$ fiber of a tensor $\mathcal X$ is any vector formed by fixing all indices of $\mathcal X$, \emph{except} the $n^{th}$ index.
    For example, a matrix column is a mode-$1$ fiber as it is defined by fixing the second index to a particular value.
    \item To indicate every element along a particular mode or dimension, we will use the $:$ symbol.
    For example, $\mathbf{A}$(1,:) denotes the first row of the matrix $\mathbf{A}$.    
    \item Tensor matricization is the process by which a tensor is \emph{unfolded} into a matrix.
    The mode-$n$ matricization of a tensor is denoted as $\mathbf X _{(n)}$, and is obtained by laying out the mode-$n$ fibers of $\mathcal X$ as the columns of $\mathbf X _{(n)}$.
    \item Khatri-Rao product (KRP)~\cite{KR_product} is the column-wise Kronecker product between two matrices, and is denoted by the symbol $\odot$. Given matrices 
	$\mathbf A^{(1)}$ $\in$ \REALTWO{I_{1}}{R}
	and $\mathbf A^{(2)}$ $\in$ \REALTWO{I_{2}}{R},
	their Khatri-Rao product \MATRIX{K},
	denoted \MATRIX{K} $=$ $\mathbf A^{(1)}$ $\odot$ $\mathbf A^{(2)}$,
	where \MATRIX{K} is a $(I_{1}\cdot I_{2}) \times R$ matrix, is defined as:	
	$\mathbf A^{(1)}\odot \mathbf A^{(2)} = \left[\textbf{a}_{1}^{(1)}\otimes \textbf{a}_{1}^{(2)}\ \textbf{a}_{2}^{(1)}\otimes \textbf{a}_{2}^{(2)} \dots \textbf{a}_{R}^{(1)}\otimes \textbf{a}_{R}^{(2)}\right] \nonumber$, where $\otimes$ denotes the Kronecker product.
    \item Element-wise product and division 
    are denoted by the symbols~$*$ and $\oslash$, respectively.
\end{enumerate}

\vspace*{-5pt}
\subsection{Tensor Decomposition}
\label{sec:tensor_rank_decomposition}
Tensor decomposition can be considered as a generalization of singular value matrix decomposition and principal component analysis, and it is used to reveal latent information embedded in large multi-dimensional data sets. 
This work targets algorithms that compute the CPD tensor decomposition model, namely, the Canonical Polyadic Alternating Least Squares (CP-ALS) algorithm for normally distributed data and the Canonical Polyadic Alternating Poisson Regression (CP-APR) algorithm for non-negative count data.

\begin{figure}[tb]
\includegraphics[width=0.99\linewidth]{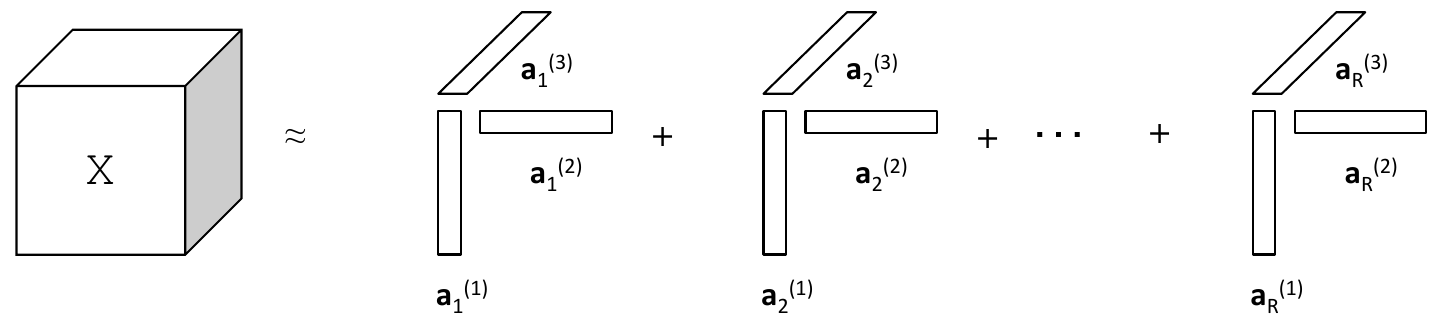}
    \vspace*{-5pt}
    \caption{CPD of a mode-$3$ tensor $\mathcal X$.
    There are $R$ rank-one tensors that are formed by the outer-product between three vectors $\mathbf a_{r}^{(1)}$, $\mathbf a_{r}^{(2)}$, and $\mathbf a_{r}^{(3)}$, where $r \in \{1,2,\ldots,R\}$.
    The vectors along the same mode are often grouped together as the columns of a \emph{factor matrix}.
    For example, the vectors $\mathbf a_{1}^{(1)}$, $\mathbf a_{2}^{(1)}$, \ldots, $\mathbf a_{R}^{(1)}$ are the columns of the mode-$1$ factor matrix $\mathbf A^{(1)}$.}
    \vspace*{-2pt}
    \label{fig:CP}
\end{figure}

CPD is a popular tensor decomposition model, where a mode-$N$ tensor \TENSOR{X} is approximated by the sum of $R$ rank-one tensors. 
A rank-one tensor is formed by $N$ vectors, each corresponding to a particular mode. 
The vectors along the same mode can be arranged as the columns of a factor matrix, resulting in $N$ different \emph{factor matrices} so that the decomposition of \TENSOR{X} is the outer product of these matrices. An example CPD of a mode-$3$ tensor is shown in Figure~\ref{fig:CP}.

\vspace*{-2pt}
\subsubsection{CP-ALS}\label{sec:cpd-als}

Algorithm~\ref{alg:cp-als} illustrates the 
CP-ALS algorithm for iteratively computing the factor matrices of the CPD model. 
In each CP-ALS iteration, every factor matrix is updated via the alternating least squares (ALS) method whereby every other factor matrix but the one being updated is fixed to yield the best approximation of \TENSOR{X}.
Line~\ref{line:mttkrp} shows the \ac{MTTKRP} operation~\cite{Smith2015}, which involves \emph{tensor matricization} and \emph{Khatri-Rao product}. 
For a mode-$3$ tensor \TENSOR{X}, the mode-1 MTTKRP operation can be expressed as
$\textbf{X}_{(1)} \left(\textbf{A}^{(2)}\odot \textbf{A}^{(3)}\right)$.
MTTKRP is typically the most expensive tensor kernel of CP-ALS, and it is performed along all modes in every CP-ALS iteration.
Since MTTKRP operations are similar across all modes, for brevity, we only discuss mode-$1$ MTTKRP in this paper.

\begin{algorithm}[tb]
\footnotesize
\caption{CP-ALS Algorithm}
\label{alg:cp-als}
\begin{algorithmic}[1]
\Require Tensor $\mathcal{X} \in \mathbb{R}^{I_{1}\times\cdots\times I_{N}}$, initial guess factor matrices $\textbf{A}^{(1)}$, $\cdots$, $\textbf{A}^{(N)}$.
\Ensure $\boldsymbol{\lambda}$,  $\textbf{A}^{(1)}$, $\cdots$, $\textbf{A}^{(N)}$
\Repeat
    \For{n = 1, $\cdots$, N}
    \State $\textbf{G}^{(1)}$ $\gets$ $\textbf{A}^{(1)T}$ $\textbf{A}^{(1)}$
    \State $\cdots$
    \State $\textbf{G}^{(n-1)}$ $\gets$ $\textbf{A}^{(n-1)T}$ $\textbf{A}^{(n-1)}$
    \State $\textbf{G}^{(n+1)}$ $\gets$ $\textbf{A}^{(n+1)T}$ $\textbf{A}^{(n+1)}$
    \State $\cdots$
    \State $\textbf{G}^{(N)}$ $\gets$ $\textbf{A}^{(N)T}$ $\textbf{A}^{(N)}$ 
    \State $\textbf{V}$ $\gets$ $\textbf{G}^{(1)}$  $*$ $\cdots$ $*$ $\textbf{G}^{(n-1)}$ $\textbf{G}^{(n+1)}$ $*$ $\cdots$ $*$ $\textbf{G}^{(N)}$
    \State $\textbf{K}$ $\gets$ $\textbf{A}^{(1)} \odot \cdots \odot \textbf{A}^{(n-1)} \odot \textbf{A}^{(n+1)} \odot \cdots \odot \textbf{A}^{(N)}$
    \State $\textbf{M}$ $\gets$ $\textbf{X}_{(n)}$ $\textbf{K}$ \Comment{MTTKRP} \label{line:mttkrp}
    \State $\textbf{A}^{(n)}$ $\gets$ $\textbf{M}$ $\textbf{V}^{\dagger}$ \Comment{Pseudoinverse}
    \State $\boldsymbol{\lambda}$ $\gets$ column normalize $\textbf{A}^{(n)}$ and store norms as $\boldsymbol{\lambda}$
    \EndFor
\Until{fit ceases to improve or maximum iterations reached}
\end{algorithmic}
\end{algorithm}

\vspace*{-2pt}
\subsubsection{CP-APR}
\label{sec:cpapr}
While CP-ALS can decompose sparse count 
data, CP-APR better describes the random variations in the data by representing it using a Poisson distribution~\cite{Chi12Tensors}, which considers a discrete number of events and assumes zero probability for observing fewer than zero events. 
As a result, CP-APR is 
more expensive to compute compared to CP-ALS~\cite{Chi12Tensors}.
There are three common methods for computing CP-APR:
\begin{inparaenum}[(i)]
    \item Multiplicative update (MU), 
    \item Projected damped Newton for row-based sub-problems (PDN-R), and
    \item Projected quasi-Newton for row-based sub-problems (PQN-R).
\end{inparaenum}

PDN-R and PQN-R employ second-order information to independently solve \emph{row sub-problems}, whereas MU uses a form of scaled steepest-descent with bound constraints over \emph{all rows} during each iteration~\cite{HaPlKo15}. Although 
MU needs more iterations to converge than PDN-R and PQN-R, it remains the most popular method due to its lower iteration cost and higher parallelism and data reuse, which makes it amenable for efficient execution on modern parallel architectures.
As such, we focus our efforts exclusively on optimizing the MU method in this study.
For a detailed discussion of the three CP-APR algorithms, we refer our readers to~\cite{HaPlKo15} and~\cite{Chi12Tensors}.

\begin{figure*}[!htbp]
    \centering
     \begin{subfigure}[b]{0.18\textwidth}
         \centering
        \begin{tabular}{ c c c | c }
            $i_{1}$ & $i_{2}$ & $i_{3}$ & $val$\\
            \hline 
            0 & 3 & 0 & 1\\ 
            1 & 0 & 0 & 2\\  
            1 & 6 & 1 & 3\\  
            2 & 2 & 1 & 4\\  
            3 & 1 & 1 & 5\\  
            3 & 4 & 0 & 6\\  
            \hline
    \end{tabular}
    \caption{COO.} \label{fig:coo}
     \end{subfigure}
     \hfill
     \begin{subfigure}[b]{0.41\textwidth}
     \centering
            \begin{tabular}{c | c c c | c c c | c }
            $b_{ptr}$ & $b_{i_{1}}$ & $b_{i_{2}}$ & $b_{i_{3}}$ & $e_{i_{1}}$ & $e_{i_{2}}$ & $e_{i_{3}}$ & $val$\\
            \hline
            0 & 0 & 1 & 0 & 0 & 1 & 0 & 1\\ 
            1 & 0 & 0 & 0 & 1 & 0 & 0 & 2\\  
            2 & 0 & 3 & 0 & 1 & 0 & 1 & 3 \\  
            3 & 1 & 1 & 0 & 0 & 0 & 1 & 4\\  
            4 & 1 & 0 & 0 & 1 & 1 & 1 & 5\\  
            5 & 1 & 2 & 0 & 1 & 0 & 0 & 6\\  
            \hline
        \end{tabular}
         \caption{HiCOO with $2\times 2\times 2$ tiling. 
         }
         \label{fig:hicoo}
     \end{subfigure}
    \begin{subfigure}[b]{0.39\textwidth}
     \centering
        \includegraphics[width=0.92\textwidth]{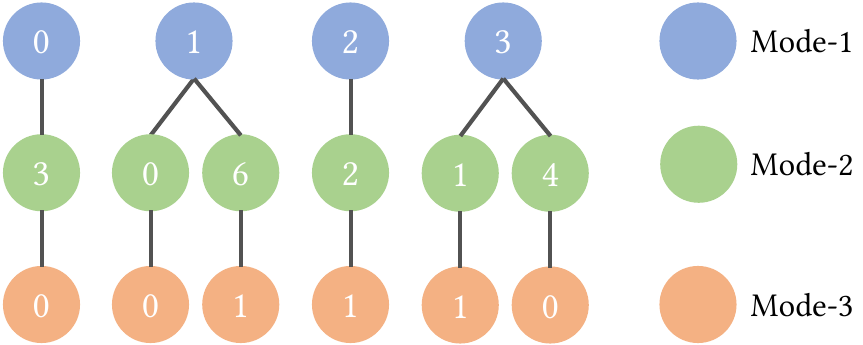}
        \caption{CSF format for mode-$1$. Each sub-tree groups the nonzero elements with the same mode-$1$ index.}
         \label{fig:csf}
     \end{subfigure} 
    \vspace*{-5pt}
    \caption{Different sparse tensor representations of a $4\times 8\times 2$ tensors with six nonzero elements.}
    \vspace*{-8pt}
    \label{fig:format-comparison}%
\end{figure*}

\begin{algorithm}[tb]
\footnotesize 
\caption{CP-APR MU Algorithm}
\label{alg:cp-apr-mu}
\begin{algorithmic}[1]
\Require Tensor $\mathcal{X} \in \mathbb{R}^{I_{1}\times\cdots\times I_{N}}$, initial guess factor matrices $\textbf{A}^{(1)}$,$\cdots$,$\textbf{A}^{(N)}$, and algorithmic parameters:
\begin{itemize}
    \item $k_{max}$, maximum number of outer iterations
    \item $l_{max}$, maximum number of inner iterations
    \item $\tau$, convergence tolerance on KKT conditions
    \item $\kappa$, inadmissible zero avoidance adjustment
    \item $\kappa_{tol}$, tolerance for potential inadmissible zero
    \item $\epsilon$, minimum divisor to prevent divide-by-zero
\end{itemize}
\Ensure $\textbf{A}^{(1)}$,$\cdots$,$\textbf{A}^{(N)}$
\For{$k=1,\ldots,k_{max}$}
    \State $\textnormal{isConverged} \gets \textnormal{true}$
    \For{$n=1,\ldots,N$} \label{line:begin_n}
        \State $\textbf{S}\left(i,r\right) \gets$
        $\begin{cases}
        \kappa, & \textnormal{if } k > 1, \textbf{A}^{(n)}(i,r) < \kappa_{tol}, \textnormal{and } \mathbf{\Phi}^{(n)}(i,r) > 1,\\
        0, & \textnormal{otherwise}
        \end{cases}$
        \State $\textbf{B} \gets \left(\textbf{A}^{(n)} + \textbf{S}\right)\mathbf{\Lambda}$
        \State $\mathbf{\Pi} \gets \left(\bigodot_{\forall m \neq n} \textbf{A}^{(m)}\right)^{T}$\Comment{Khatri-Rao product (KRP)}  \label{line:pi}
        \For{$l=1,\ldots,l_{max}$}
            \State $\mathbf{\Phi}^{(n)} \gets \left(\mathbf{X}_{(n)} \oslash  \left(\max\left(\textbf{B}\mathbf{\Pi},\epsilon\right)\right) \right)\mathbf{\Pi}^{T}$ \label{line:phi}
            \If{$\lvert\min(\textbf{B}, \textbf{E} - \mathbf{\Phi}^{(n)})\rvert < \tau$} \Comment{Convergence check}\label{line:convergence}
                \State break
            \EndIf
            \State $\textnormal{isConverged} \gets \textnormal{false}$
            \State $\textbf{B} \gets \textbf{B} * \mathbf{\Phi}^{(n)}$ \label{line:mu}
        \EndFor
        \State $\lambda \gets e^{T}\textbf{B}$, ~~$\textbf{A}^{(n)} \gets \textbf{B}\mathbf{\Lambda}^{-1}$
    \EndFor
    \If{isConverged $=$ true}
        \State break
    \EndIf
\EndFor
\end{algorithmic}
\end{algorithm}
Algorithm~\ref{alg:cp-apr-mu} shows CP-APR with the MU method. 
Using two nested loops, CP-APR computes the decomposition of a tensor by successively updating each factor matrix while holding the other factor matrices fixed. 
Lines~\ref{line:pi} and~\ref{line:phi} show the $\mathbf{\Pi}$ (Khatri-Rao product) and $\mathbf{\Phi}$ (model update) kernels, respectively, which make up majority of the CP-APR execution time.
Note that $\mathbf{\Pi}$ is calculated once per \emph{outer} loop \emph{for each mode}, and $\mathbf{\Phi}$ is calculated once per \emph{inner} loop. Since the inner loop is executed at most $l_{max}$ times \emph{per mode per outer loop}, the cost of calculating $\mathbf{\Pi}$ can more easily be amortized as the number of inner iterations goes up.

\vspace*{-5pt}
\subsection{Sparse Tensor Storage Formats}
\label{subsec:formats}
We present an overview of raw and compressed sparse tensor storage using three popular formats: coordinate (COO), hierarchical coordinate (HiCOO), and compressed sparse fiber (CSF).
Figure~\ref{fig:format-comparison} shows a comparison of the different formats for a $4\times 8\times 2$ tensor with six nonzero elements.

\vspace*{-5pt}
\subsubsection{Coordinate (COO)}
COO is the canonical and simplest sparse format, as it \emph{lists} the nonzero elements and their 
$N$-dimensional coordinates, without any compression. 
This mode-agnostic form allows tensor algorithms to use one tensor copy across modes.
Figure~\ref{fig:format-comparison}(a) shows an example 
tensor in the COO format.  

Decomposing a sparse tensor stored in the COO format typically involves iterating over each nonzero element 
and updating the corresponding factor matrix row.
For example, during mode-$1$ computation, a nonzero element with coordinates ($i_{1}$,$i_{2}$,$i_{3}$) updates row $i_{1}$ of the mode-$1$ factor matrix after reading rows $i_{2}$ and $i_{3}$ from mode-$2$ and mode-$3$ factor matrices, respectively.
Since multiple threads can simultaneously update the same row 
of 
a factor matrix, 
these updates must be done atomically, which can be expensive on parallel processors with a large number of threads.

\vspace*{-5pt}
\subsubsection{Hierarchical Coordinate (HiCOO)}
HiCOO~\cite{li2018hicoo,li2019efficient} is a block-based sparse tensor format that employs multi-dimensional tiling for data compression.
Like COO, HiCOO is mode-agnostic but its compression efficacy  depends completely on the properties of the target tensor, such as its shape, density, and data distribution, and determining the optimal parameters for compression (e.g., the tile size) is non-trivial.
In some cases, rearranging the nonzero elements to create dense tiles is necessary to achieve any compression~\cite{li2019efficient}.
In addition, scheduling the resulting HiCOO blocks can suffer from limited parallelism, due to conflicting updates across blocks, as well as workload imbalance if some blocks have significantly more nonzero elements than others.
Figure~\ref{fig:format-comparison}(b) shows the example sparse tensor encoded in the HiCOO format.
The memory required to keep the hierarchical indices (i.e., $b_{i_{1}}$, $e_{i_{1}}$, etc.) can be lower than storing the actual indices (i.e., $i_{1}$, $i_{2}$, and $i_{3}$) only if each tile has a sufficient number of nonzero elements.

\vspace*{-5pt}
\subsubsection{Compressed Sparse Fiber (CSF)}
CSF stores a tensor as a collection of sub-trees, where each sub-tree represents a group of \emph{all} nonzero elements that update the same factor matrix row. 
Given a CSF representation with a mode ordering of 1-2-3, where 1 is the root mode and 3 is the leaf mode, the root nodes represent the rows that will be updated, and the leaf nodes represent the nonzero elements that contribute to that update.
Thus, iterating over the nonzero elements 
involves a bottom-up traversal of each sub-tree, such that at each non-leaf node, the partial results from its children are merged and pushed up, and this propagates until results from every node in the tree are merged at the root. Figure~\ref{fig:format-comparison}(c) illustrates the CSF sub-trees created from the example sparse tensor.

One advantage of CSF is that updates to 
factor matrices can be done asynchronously 
by assigning one thread to each sub-tree.
However, CSF requires $N$ tensor copies to maintain synchronization-free updates across every mode, which can be impractical for large tensors and/or devices with limited memory capacity (e.g., GPUs).
Alternatively, the sub-trees can be traversed both bottom-up and top-down, merging partial results at the tree level corresponding to the destination mode. 
While this approach allows a single tensor copy 
(with the root mode chosen arbitrarily) to be used across all modes, it requires  synchronization to avoid update conflicts and
entirely different tree traversal algorithms~\cite{Smith2015a}.
Additionally, regardless of the strategy used, CSF suffers from workload imbalance, as some sub-trees can have significantly more nonzero elements than the others.

%% file: text/approach.tex
\vspace*{-5pt}
\section{ALTO Format}\label{sec:approach}
To tackle the highly irregular shapes and distributions of real-world sparse data, our \ALTO format maps the coordinates of a nonzero element in the $N$-dimensional space that represents a tensor to a mode-agnostic index in a compact \emph{linear space}. Specifically, \ALTO uses a data-aware recursive encoding to partition every mode of the original Cartesian space into multiple regions 
such that each distinct mode has a variable number of regions to adapt to the unequal cardinalities of different modes and to minimize the 
storage requirements.
This adaptive linearization and recursive partitioning of the multi-dimensional space ensures that neighboring points in space are close to each other on the resulting compact line, thereby maintaining the inherent data locality of tensor algorithms. 
Moreover, the \ALTO format is not only locality-friendly, but also parallelism-friendly as it allows partitioning of the multi-dimensional space into perfectly balanced (in terms of workload) subspaces.
Further, it intelligently arranges the modes in the derived subspaces based on their cardinality (dimension length) to further reduce the overhead of resolving the update conflicts that typically occur in parallel sparse tensor computations.

What follows is a detailed description and discussion 
of the \ALTO format generation ($\S$\ref{sec:approach-gen}) 
using a walk-through example. Additionally, we present the \ALTO-based sequential algorithm for the MTTKRP operation
($\S$\ref{sec:approach-mttkrp}). 

\vspace*{-5pt}
\subsection{ALTO Tensor Generation}\label{sec:approach-gen}
Formally, an \ALTO tensor \TENSOR{X} $= \{$\VECELEM{x}{1}, \VECELEM{x}{2}, $\dots$, \VECELEM{x}{M}$\} $ is an ordered set of nonzero elements, where each element \VECELEM{x}{i} $= \langle v_{i}, p_{i} \rangle $ is represented by a value $v_{i}$ and a position $p_{i}$. 
The position $p_{i}$ corresponds to a compact mode-agnostic encoding of the indexing metadata, which is used to 
quickly generate the tuple $(i_{1},i_{2},\dots,i_{N})$ that locates a nonzero element 
in the multi-dimensional Cartesian space.

The generation of an \ALTO tensor 
is carried out in two stages: linearization and ordering. First, \ALTO constructs the indexing metadata using a compressed encoding scheme, based on the cardinalities of tensor modes, to map each nonzero element to a position on a compact line. Second, it arranges the nonzero elements in linearized storage according to their line positions, i.e., the values of their \ALTO index. 
Typically, the ordering stage dominates the format generation time. However, compared to the other compressed sparse tensor formats~\cite{Smith2015, Smith2015a, li2018hicoo, li2019efficient, nisa2019load, nisa2019efficient}, \ALTO requires a minimal generation time because ordering the linearized tensors incurs 
a fraction of the cost required to sort multi-dimensional index sets (due to the reduction in comparison operations, as detailed in $\S$\ref{sec:perf}).

\begin{figure}[tb]
\centering
\includegraphics[width=0.99\linewidth]{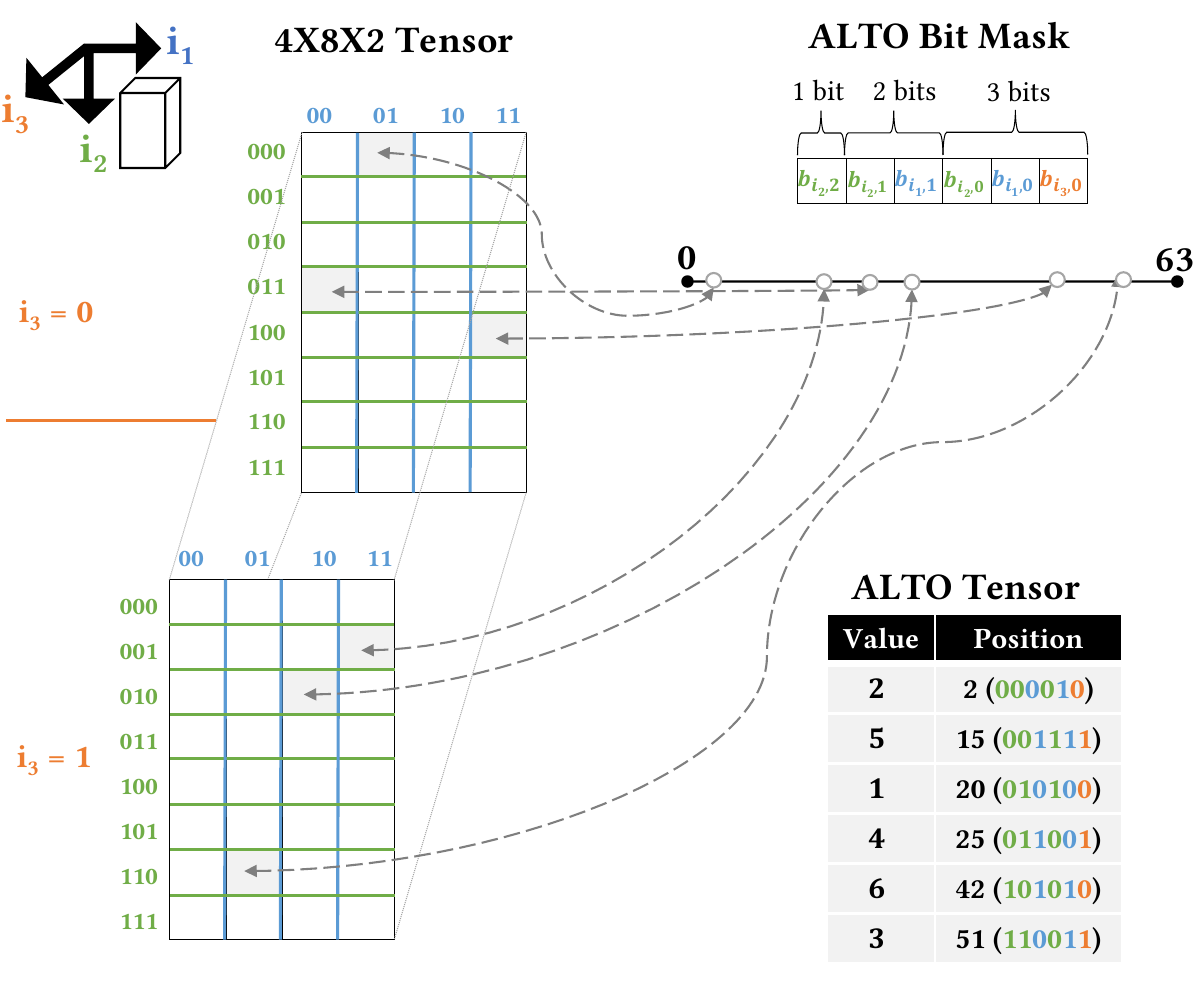}
	\vspace*{-12pt}
	\caption{An example of the \ALTO sparse encoding and representation for the three-dimensional tensor in Figure~\ref{fig:format-comparison}(a).}
	\vspace*{-10pt}
\label{fig:alto_encoding}
\end{figure}

Figure~\ref{fig:alto_encoding} shows the \ALTO format for the sparse tensor from Figure~\ref{fig:format-comparison}(a).
The multi-dimensional indices ($i_{1}$, $i_{2}$, and $i_{3}$) are color coded and the $r^{th}$ bit of their binary representation is denoted $b_{i_{n}, r}$. 
Specifically, \ALTO keeps the value of a nonzero element along with a linearized index, where each bit of this index is selected to partition the multi-dimensional space into two hyperplanes.
For example, the \ALTO encoding in Figure~\ref{fig:alto_encoding} uses a compact line of length $64$ (i.e., a $6$-bit linearized index) to represent the target tensor of size $4\times8\times2$. This index consists of three groups of bits with variable sizes (resolutions) to efficiently handle high-order data of arbitrary dimensionality. Within each bit group, \ALTO arranges the modes in an increasing order of their length (i.e., the shortest mode first), which is equivalent to partitioning the multi-dimensional space along the longest mode first. 
Such an encoding aims to generate a balanced linearization of irregular Cartesian spaces, where 
the position resolution of a nonzero element increases with every consecutive bit, starting from the most significant bit. Therefore, the line segments encode subspaces with mode intervals 
of equivalent lengths, e.g., the line segments $[0-31]$, $[0-15]$, and $[0-7]$ encode subspaces of $4\times4\times2$, $4\times2\times2$, and $2\times2\times2$ dimensions, respectively.   

Due to this adaptive encoding, \ALTO represents the resulting linearized index using the minimum number of bits, 
and it improves data locality across all modes of a given sparse tensor. Hence, a mode-$N$ tensor, whose dimensions are $I_{1}\times I_{2}\times \cdots \times I_{N}$, can be efficiently represented using a single \ALTO format with indexing metadata of size:
\vspace*{-4pt}
\begin{equation}
    S_{\text{ALTO}} = M \times (\sum_{n=1}^{N} \log_2 I_{n}) \text{ bits,}
\end{equation}
where $M$ is the number of nonzero elements.

As a result, compared to the de facto COO format, 
\ALTO reduces the storage requirement regardless of the tensor's characteristics. That is, the metadata compression ratio of the \ALTO format relative to COO is always $\ge 1$. On a hardware architecture 
with a word-level memory addressing mode, this compression ratio is given by:
\vspace*{-4pt}
\begin{equation}
    \frac{S_{\text{COO}}}{S_{\text{ALTO}}} = \frac{\sum_{n=1}^{N} \Big\lceil\frac{\log_2 I_{n}}{W_b}\Big\rceil}{\bigg\lceil\frac{\sum_{n=1}^{N} \log_2 I_{n}}{W_b}\bigg\rceil},
\end{equation}
where $W_b$ is the word size in bits. 
For example, 
on an architecture with byte-level addressing
(i.e., $W_b = 8$ bits), the sparse tensor in Figure~\ref{fig:alto_encoding} needs three bytes to store the mode indices 
(coordinates) for each nonzero element in the 
COO format, whereas only a single byte is required to store the linearized index 
in the ALTO format: the metadata compression ratio of \ALTO compared to 
COO is \emph{three}.

\begin{figure}[tb]
\centering
\vspace*{-5pt}
\includegraphics[width=1.0\linewidth]{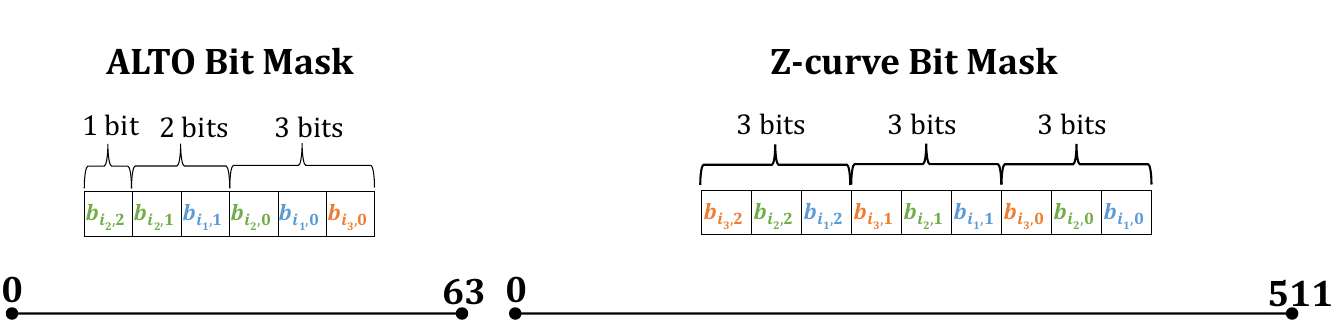}
	\vspace*{-18pt}
	\caption{For the example in Figure~\ref{fig:alto_encoding}, \ALTO generates a non-fractal, yet more compact encoding compared to traditional space-filling curves, such as the Z-Morton order.}
\label{fig:alto_sfc}
\end{figure}

\begin{figure}[tb]
\centering
\subfloat[\ALTO generates its 
linearized index using bit-level gather operations.]{
\includegraphics[width=1.0\linewidth]{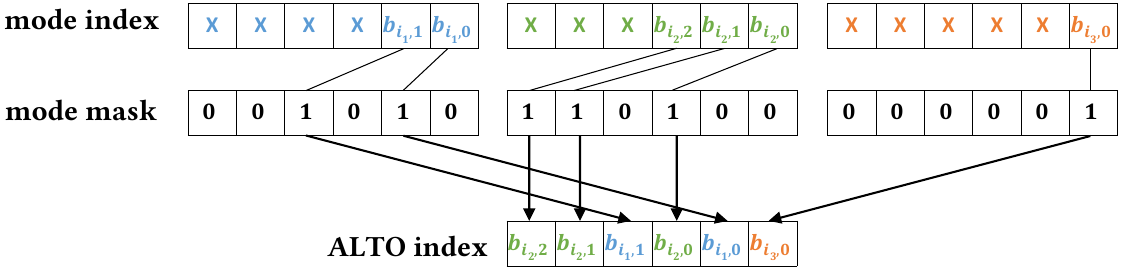}
}\vspace*{4pt}

\subfloat[To recover the multi-dimensional indices, \ALTO decodes the linearized indexing metadata using bit-level scatter operations.]{
\includegraphics[width=1.0\linewidth]{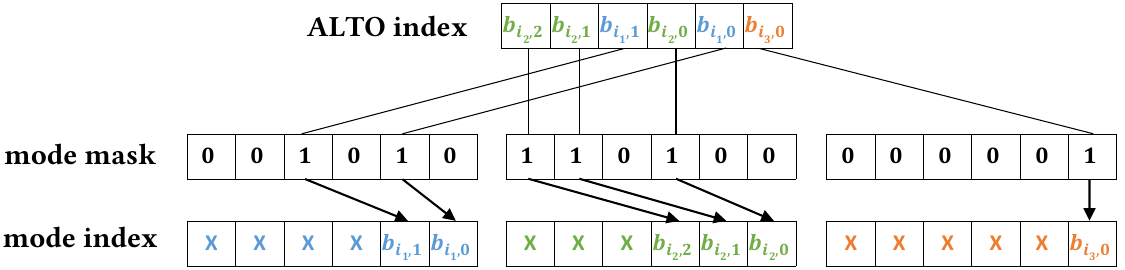}
}
	\vspace*{-5pt}
\caption{The \ALTO-based bit encoding and decoding mechanisms for the example in Figure~\ref{fig:alto_encoding}.}
	\vspace*{-7pt}
\label{fig:alto_linear}
\end{figure}

Moreover, the \ALTO format not only reduces the 
memory traffic of sparse tensor computations, but also decreases the number of memory transactions required to access the indexing metadata of a sparse tensor, as reading the linearized index requires fewer 
accesses compared to reading several multi-dimensional indices. In addition, this natural coalescing of the multi-dimensional indices into a single linearized index increases the memory transaction size to make more efficient use of the main memory bandwidth.  

It is important to note that \ALTO uses a non-fractal\footnote{A fractal pattern is a hierarchically self-similar pattern that looks the same at increasingly smaller scales.} encoding scheme, unlike the traditional space-filling curves (SFCs)~\cite{peano1890courbe}. In contrast, SFCs (e.g., Z-Morton order~\cite{morton1966computer}) are based on continuous self-similar (or fractal) functions that target dense data, which can be extremely inefficient when used to encode the irregularly shaped multi-dimensional spaces that emerge in higher-order sparse tensor algorithms as they require indexing metadata of size:
\vspace*{-4pt}
\begin{equation}
    S_{\text{SFC}} = M \times \left(N \times \max_{n=1}^{N} (\log_2 I_{n})\right) \text{ bits.}
\end{equation}

Therefore, 
in sparse tensor computations, SFCs have been \textit{only used to reorder the nonzero elements to improve their 
data locality} rather than compressing the indexing metadata~\cite{li2018hicoo}. Figure~\ref{fig:alto_sfc} shows the compact encoding generated by \ALTO compared to the fractal encoding scheme of the Z-Morton curve. In this example, 
\ALTO reduces the length of the encoding line by a factor of eight, which not only decreases the overall size of the indexing metadata, but also reduces the linearization/de-linearization time required to map the multi-dimensional space to/from the 
encoding line.

To allow fast indexing of linearized tensors during sparse tensor computations, the \ALTO encoding is implemented using a set of simple $N$ bit masks, where $N$ is the number of modes, on top of common data processing primitives. Figure~\ref{fig:alto_linear} shows the linearization and de-linearization mechanisms, which are used during the \ALTO format generation and the sparse tensor computations, respectively. The linearization process is implemented as a bit-level gather, while the de-linearization is performed as a bit-level scatter. Thus, although the compressed representation of the proposed \ALTO format comes at the cost of a de-linearization (decompression) overhead, such a computational overhead is minimal and can be effectively overlapped with the memory accesses of the memory-intensive sparse tensor operations, as shown in $\S$\ref{sec:perf}. 

\subsection{ALTO-based Tensor Operations}\label{sec:approach-mttkrp}

Since 
high-dimensional data analytics is becoming increasingly popular in rapidly evolving areas~\cite{ho2014marble, papalexakis2016tensors, sidiropoulos2017tensor, kobayashi2018extracting}, a fundamental goal of the proposed \ALTO format is to deliver superior performance without compromising the productivity of end users to allow fast development of tensor algorithms.
To this end, Algorithm~\ref{fig:mttkrp-algo} illustrates 
the popular MTTKRP tensor operation using the \ALTO format. 

First, unlike mode-specific (e.g., CSF-based) formats, \ALTO enables end users to perform tensor operations using \textit{a unified code implementation that works on a single tensor copy} 
regardless of the different mode orientations of such operations. Second, by decoupling the representation of a sparse tensor from the distribution of its nonzero elements, \ALTO does not require 
\textit{manual} tuning to select the optimal \textit{format parameters} for this tensor, in contrast to prior approaches such as HiCOO and CSF.
Instead, the \ALTO format is automatically generated based on the shape and dimensions of the target sparse tensor (as explained in $\S$\ref{sec:approach-gen}).

\begin{algorithm}[htb]
\footnotesize 
\begin{algorithmic}[1]
\Require A third-order \ALTO sparse tensor \TENSOR{X} $\in$ \REALTHREE{I_{1}}{I_{2}}{I_{3}} with $M$ nonzero elements, dense factor matrices $\textbf{A}^{(1)}$,$\textbf{A}^{(2)}$,$\textbf{A}^{(3)}$ 
\Ensure Updated dense factor matrix \MATRIX{\~{A}} $\in$ \REALTWO{I_{1}}{R}
\For{ $x = 1, \dots, M$}
\State $\textbf{i} = \mathbb{EXTRACT}(pos(x), MASK)$ \Comment{De-linearization.}
    \State \MATRIX{\~{A}}$(i_{1},:)$ $+= val(x) \times$ $\textbf{A}^{(2)}$$(i_{2},:) \times$ $\textbf{A}^{(3)}$$(i_{3},:)$
\EndFor
\Return \MATRIX{\~{A}}
\end{algorithmic}
\caption{Mode-$1$ sequential MTTKRP-\ALTO algorithm.}
\label{fig:mttkrp-algo}
\end{algorithm}

\section{Parallel Linearized Tensor Algorithms}\label{sec:approach-algo}
We devise a set of \ALTO-based parallel algorithms for accelerating sparse tensor computations and demonstrate how they are employed in popular TD operations.  

\subsection{Workload Partitioning and Scheduling}\label{sec:approach-par}
Prior compressed sparse tensor formats partition
the tensor space into \emph{non-overlapping} regions and cluster the data into coarse-grained structures, such as blocks, slices, and/or fibers.
Due to the irregular shapes and distributions of higher-order data, such coarse-grained approaches can suffer from severe workload imbalance
and limited parallel performance/scalability.  
In contrast, by employing the \ALTO format, linearized tensor algorithms work
at the finest granularity level (i.e., a single nonzero element), which allows perfect load-balancing and  scalable parallel execution. 
While a non-overlapping partitioning 
can be obtained from the \ALTO encoding scheme by using a subset of the index bits, the workload balance of such a partitioning still depends on the sparsity patterns of the tensor. 

\begin{figure}[tb]
\centering
\includegraphics[width=1.0\linewidth]{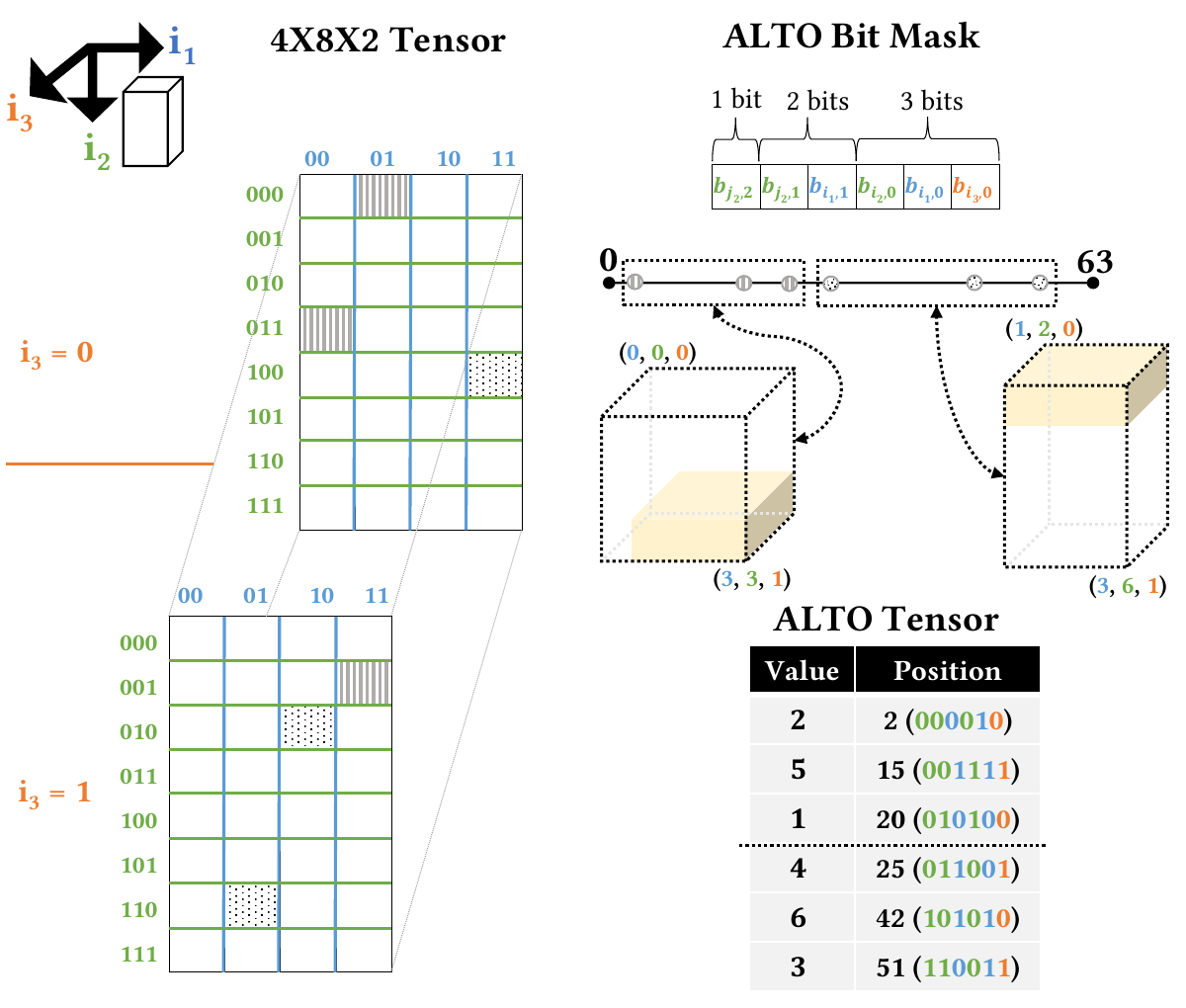}
	\vspace*{-15pt}
	\caption{\ALTO partitioning of the example in Figure~\ref{fig:alto_encoding},
	which generates balanced partitions
	in terms of workload (nonzero elements) for efficient parallel execution.
	}
 \vspace*{-5pt}
\label{fig:alto_parallel}
\end{figure}

To decouple the performance of sparse tensor computations from the distribution of nonzero elements, we divide the multi-dimensional space into potentially \emph{overlapping} regions and allow workload distribution at the granularity of nonzero elements, which result in perfectly balanced partitions in terms of workload.
Figure~\ref{fig:alto_parallel} depicts an example of \ALTO's workload decomposition when applied to the sparse tensor in Figure~\ref{fig:alto_encoding}.
\ALTO divides the line segment containing the nonzero elements of the target tensor into two smaller line segments: $[2-20]$ and $[25-51]$, which have different lengths (i.e., 18 and 26) but the same number of nonzero elements, thus perfectly splitting the workload.

Once the linearized sparse tensor is divided into multiple line segments, \ALTO identifies the basis mode intervals (coordinate ranges) of the multi-dimensional subspaces that correspond to these segments. For example, the line segments $[2-20]$ and $[25-51]$ correspond to three-dimensional subspaces bounded by the mode intervals $\{[0-3], [0-3], [0-1]\}$ and $\{[1-3], [2-6], [0-1]\}$, respectively. While the derived multi-dimensional subspaces of the line segments may overlap, as highlighted in yellow in Figure~\ref{fig:alto_parallel}, each nonzero element is assigned to exactly one line segment. That is, \ALTO imposes a partitioning on a given linearized tensor that generates a disjoint set of \textit{non-overlapping and balanced line segments}, but it does not guarantee that such a partitioning will divide the multi-dimensional space of the tensor into non-overlapping subspaces. In contrast, the prior compressed formats split the multi-dimensional space into non-overlapping (yet highly imbalanced) regions, namely, tensor slices and fibers in CSF-based formats and multi-dimensional blocks in block-based formats (e.g., HiCOO).

Formally, a set of $L$ line segments partitions an 
\ALTO tensor \TENSOR{X}, which encodes $N$-dimensional sparse data,
such that \(\TENSORI{X} = \TENSORI{X}_1 \cup \TENSORI{X}_2 \cdots \cup \TENSORI{X}_L\) and
\(\TENSORI{X}_{i} \cap \TENSORI{X}_{j} = \phi \forall i~\text{and}~j\), where $i \neq j$.
Each line segment \TENSOR{X}$_{l}$ is an ordered set of nonzero elements that are bounded in an $N$-dimensional space by a set of $N$ closed mode intervals
\(T_{l}~=~\{ [T_{l,1}^{s}, T_{l,1}^{e}], [T_{l,2}^{s}, T_{l,2}^{e}], \cdots [T_{l,N}^{s}, T_{l,N}^{e}]\}\),
where each mode interval $T_{l,n}$ is delineated by a start $T_{l,n}^{s}$ and an end $T_{l,n}^{e}$. The intersection of 
two mode interval sets represents the subspace overlap between their corresponding line segments (partitions), as highlighted in yellow in Figure~\ref{fig:alto_parallel}.
This overlap information is used to more efficiently resolve conflicts between partitions, as described in~$\S$\ref{sec:approach-cf}.

\subsection{Adaptive Conflict Resolution}\label{sec:approach-cf}
\begin{algorithm}[tb]
\
\newcommand{\algcolor}[2]{\hspace*{-\fboxsep}\colorbox{#1}{\parbox{\dimexpr\linewidth-\fboxsep}{#2}}}
\newcommand{\algemph}[1]{\algcolor{lighter-gray}{#1}}
\newcommand{\algemphe}[1]{\algcolor{light-gray}{#1}}
\begin{algorithmic}[1]
\footnotesize 
\Require A third-order \ALTO sparse tensor \TENSOR{X} $\in$ \REALTHREE{I_{1}}{I_{2}}{I_{3}} with $M$ nonzero elements, dense factor matrices $\textbf{A}^{(1)}$,$\textbf{A}^{(2)}$,$\textbf{A}^{(3)}$ 
\Ensure Updated dense factor matrix \MATRIX{\~{A}} $\in$ \REALTWO{I_{1}}{R}
\For{ $l = 1, \dots, L$ \textbf{in parallel}} \Comment{\ALTO line segments.}
    \For{ $\forall x \in$ \TENSOR{X}$_{l}$}
    \State \colorbox{lighter-gray}{$p = pos(x)$~~$v = val(x)$}
    \State \colorbox{light-gray}{$p = pos\_out(x)$~~$v = val\_out(x)$}\vspace*{\fboxsep}    
    \State $ \textbf{i} = \mathbb{EXTRACT}(p, MASK(1))$ \Comment{De-linearization.}
        \State \colorbox{lighter-gray}{\MATRIX{Temp}$_{l}(i_{1}-T_{l,1}^{s},:)$ $+=v 
 \times$$\textbf{A}^{(2)}$$(i_{2},:) \times$$\textbf{A}^{(3)}$$(i_{3},:)$}
        \algemphe{
        \If{$i_{1}$ \textit{is} boundary $=$ true}
        \State $\mathbb{ATOMIC}($\MATRIX{\~{A}}$(i_{1},:)$ $+=v \times$$\textbf{A}^{(2)}$$(i_{2},:) \times$$\textbf{A}^{(3)}$$(i_{3},:)$)
        \Else
        \State \MATRIX{\~{A}}$(i_{1},:)$ $+= v \times$$\textbf{A}^{(2)}$$(i_{2},:) \times$$\textbf{A}^{(3)}$$(i_{3},:)$
        \EndIf
        }
    \EndFor
\EndFor
\item[]
\algemph{
\For{ $b = 1, \dots, I_{1}$ \textbf{in parallel}} \Comment{Pull-based reduction.}
    \For{ $\forall l$  where $b \in [T_{l,1}^{s}, T_{l,1}^{e}]$}
        \State \MATRIX{\~{A}}$(b,:)$ $+=$ \MATRIX{Temp}$_{l}(b-T_{l,1}^{s},:)$
    \EndFor
\EndFor
}
\Return \MATRIX{\~{A}}
\end{algorithmic}
\caption{
Parallel mode-$1$ MTTKRP-\ALTO algorithm. \ALTO 
automatically uses either \colorbox{lighter-gray}{recursive} or \colorbox{light-gray}{output-oriented} tensor traversal, based on the reuse of output fibers, 
to efficiently resolve update conflicts. }
\label{fig:mttkrp-par-algo}
\end{algorithm}

Because processing the nonzero elements of a tensor in parallel (e.g., line~1 in Algorithm~\ref{fig:mttkrp-algo}) can result in write conflicts across threads (e.g., line~3 in Algorithm~\ref{fig:mttkrp-algo}),
we devise an adaptive parallel 
algorithm to handle these conflicts by exploiting the inherent data reuse of target tensors. 
Our adaptive algorithm chooses between 
    1) recursively traversing the tensor 
    to maximize the reuse of both the input and output fibers, at the cost of global parallel reduction, or
    2) traversing the tensor elements in output-oriented ordering, and only synchronizing at partition boundaries.

Algorithm~\ref{fig:mttkrp-par-algo} describes the proposed parallel execution scheme using a representative MTTKRP operation that works on a sparse tensor stored in the \ALTO format. After \ALTO imposes a partitioning on a sparse tensor, as detailed in $\S$\ref{sec:approach-par}, each partition can be assigned to a different thread. 
To resolve the update/write conflicts that may happen during parallel sparse tensor computations,
\ALTO uses an \textit{adaptive conflict resolution} approach that \textit{automatically} selects the appropriate tensor traversal and 
synchronization technique (highlighted by the different gray backgrounds) 
based on the reuse of the target fibers. This metric represents the average number of nonzero elements per fiber (i.e., the generalization of a matrix row or column) and it is computed in constant time by simply dividing the total number of nonzero elements by the number of fibers along the target mode. 

\begin{figure}[tb]
\centering
\includegraphics[width=1.0\linewidth]{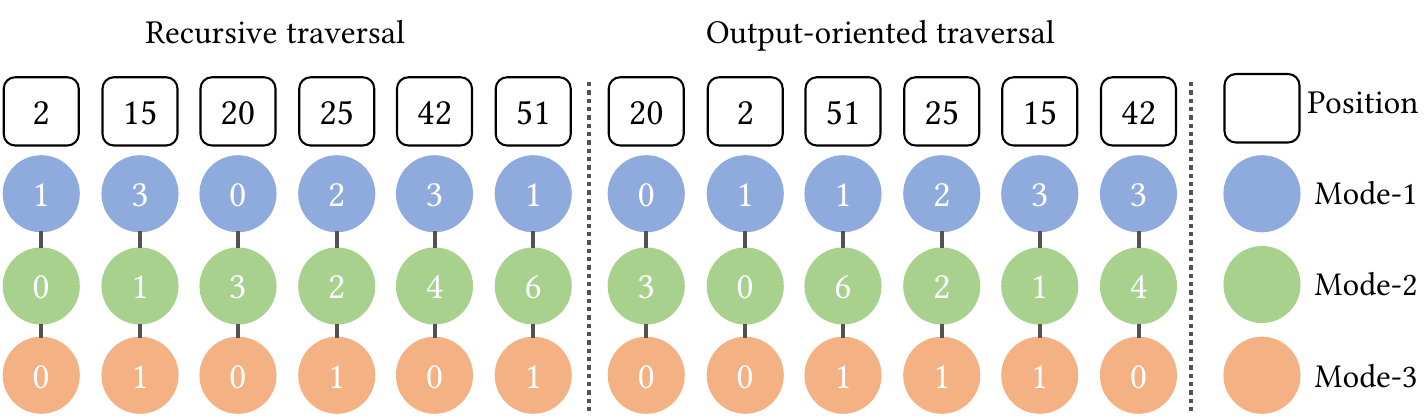}
	\vspace*{-15pt}
	\caption{Recursive vs. output-oriented traversal of the example tensor in Figure~\ref{fig:alto_encoding}, where mode-$1$ is the target/output mode. The coordinates of each nonzero (mode indices) are extracted from its line position (linearized index) as detailed in Figure~\ref{fig:alto_linear}.
	}
 \vspace*{-10pt}
\label{fig:alto_out}
\end{figure}

When a sparse tensor operation exhibits high fiber reuse, \ALTO recursively traverses the tensor (by accessing the nonzero elements in an increasing order of their linearized index or line position as illustrated in Figure~\ref{fig:alto_out}) and it uses a limited amount of temporary (local) storage to capture the local updates of different partitions (line~6). Next, \ALTO merges the conflicting global updates (lines~14--18) using an efficient pull-based parallel reduction, where the final results are computed by pulling the partial results from 
the \ALTO partitions. When computing the factor matrix of the target mode (e.g., mode-$1$), such a recursive tensor traversal 1) increases the likelihood that both input (mode-$2$/$3$)
and output (mode-$1$) fibers remain in fast memories, and 2) reduces the size of temporary storage (partial copy of mode-$1$ factor matrix) needed for each partition, which in turn decreases the overhead of the pull-based reduction.

\ALTO considers the fiber reuse large enough to 
use local staging memory for conflict resolution, if the average number of nonzero elements per fiber is \textit{more than} the maximum cost of using this two-stage (buffered) accumulation process, which consists of initialization (omitted for brevity), local accumulation (line~6 in Algorithm~\ref{fig:mttkrp-par-algo}), and global accumulation (lines~14--18). In the worst (no reuse) case, the buffered accumulation cost is four memory operations (two read and two write operations). 
As explained in $\S$\ref{sec:approach-par}, each line segment \TENSOR{X}$_{l}$ is bounded in an $N$-dimensional space by a set of $N$ closed mode intervals $T_{l}$, which is computed during the partitioning of an \ALTO tensor; thus, the size of the temporary storage accessed during the accumulation of \TENSOR{X}$_{l}$'s updates along a mode $n$ is directly determined by the mode interval $[T_{l,n}^{s}, T_{l,n}^{e}]$ (see lines~6 and 15).   

When the target tensor computations suffer from limited fiber reuse, \ALTO uses output-oriented tensor traversal, where the nonzero elements are accessed in an increasing order of their target/output mode (e.g. mode-$1$ as depicted in Figure~\ref{fig:alto_out}). That way, the data reuse of output fibers is fully captured and synchronization across threads can be avoided. Specifically, \ALTO needs to resolve the conflicting updates across its line segments (partitions) using direct atomic operations (line~8) only if the output fiber is at the boundary between different partitions/threads. This output-oriented traversal resembles the CSF-based tree traversal (see Figure~\ref{fig:format-comparison}(c)); however, in contrast to CSF, \ALTO uses a fine-grained compact index (line position) to encode nonzero coordinates instead of a coarse-grained index tree, which requires a single tensor copy (instead of one copy per mode) and allows perfect load balancing during parallel execution. In addition, our output-oriented tensor traversal is only used when fiber reuse is limited; otherwise, the recursive traversal method is employed because of its superior data locality and parallel performance, thanks to reusing both input and output fibers as well as amortizing the overhead of synchronization operations (pull-based reduction).

\subsection{Adaptive Memory Management}\label{subsec:phi}

\begin{algorithm}[tb]
\footnotesize
\newcommand{\algcolor}[2]{\hspace*{-\fboxsep}\colorbox{#1}{\parbox{\dimexpr\linewidth-\fboxsep}{#2}}}
\newcommand{\algemph}[1]{\algcolor{lighter-gray}{#1}}
\newcommand{\algemphe}[1]{\algcolor{light-gray}{#1}}
    \begin{algorithmic}[1]
        \For{ $l = 1, \dots, L$ \textbf{in parallel}} \Comment{ALTO line segments} \label{line:par}
            \For{ $\forall x \in$ \TENSOR{X}$_{l}$}
            \State \colorbox{lighter-gray}{$p = pos(x)$~~$v = val(x)$}
            \State \colorbox{light-gray}{$p = pos\_out(x)$~~$v = val\_out(x)$}\vspace*{\fboxsep}   
                \State $\textbf{i} = \mathbb{EXTRACT}(p, MASK)$ \Comment{Delinearization} \label{line:delin} 
                \If{preCompute $\mathbf{\Pi}$ = true}
                    \State $\textbf{krp} \gets \mathbf{\Pi}(x,:)$\label{line:pi-precompute}
                \Else
                \State $\textbf{krp} \gets \left(*_{\forall m \neq n} \textbf{A}^{(m)}(i_{m},:)\right)$\label{line:otf-krp}
                \EndIf
                \State \colorbox{lighter-gray}{\MATRIX{Temp}$_{l}(i_{n}-T_{l,n}^{s},:)$ $\mathrel{+}=$ $\left(\frac{val(x)}{\max\left(\textbf{B}(i_{n},:)\textbf{krp}^{T},\epsilon\right)}\right)\textbf{krp}$} \label{line:sm}
        \algemphe{
        \If{$i_{n}$ \textit{is} boundary $=$ true}
        \State $\mathbb{ATOMIC}\left(\mathbf{\Phi^{(n)}}(i_{n},:) \mathrel{+}= \left(\frac{v}{\max\left(\textbf{B}(i_{n},:)\textbf{krp}^{T},\epsilon\right)}\right)\textbf{krp}   \right)$\label{line:atomic}
        \Else
        \State $\mathbf{\Phi^{(n)}}(i_{n},:) \mathrel{+}= \left(\frac{v}{\max\left(\textbf{B}(i_{n},:)\textbf{krp}^{T},\epsilon\right)}\right)\textbf{krp}$
        \EndIf
        }
            \EndFor
        \EndFor
         \item[]
        \algemph{
        \For{ $b = 1, \dots, I_{n}$ \textbf{in parallel}} \Comment{Pull-based reduction} \label{line:reduce-begin} 
            \For{ $\forall l$  where $b \in [T_{l,n}^{s}, T_{l,n}^{e}]$}
                \State $\textbf{$\Phi^{(n)}$}(b,:)$ $+=$ \MATRIX{Temp}$_{l}(b-T_{l,n}^{s},:)$
            \EndFor
        \EndFor \label{line:reduce-end} 
        }
        \Return \textbf{$\Phi^{(n)}$}
    \end{algorithmic}
    \caption{
    Parallel mode-$1$ $\mathbf{\Phi}$-\ALTO kernel. ALTO 
    performs \emph{either} \colorbox{lighter-gray}{recursive} or \colorbox{light-gray}{output-oriented} tensor traversal, based on fiber reuse,
    to efficiently resolve update conflicts. In addition, it determines whether to 
    reuse or recompute intermediate results.   
    }
    \label{fig:subproblem_algo}
\end{algorithm}

In many tensor decomposition 
algorithms, such as CP-APR, the intermediate results of tensor kernels are typically stored and then reused during the iterative optimization loop (as shown in Algorithm~\ref{alg:cp-apr-mu}). However, storing these important calculations can substantially increase memory traffic and require prohibitive amount of memory, especially for large tensors and high decomposition ranks. 
Hence, in contrast to the traditional algorithm that pre-computes and reuses the intermediate values (\ALTO-PRE), we introduce an \ALTO-based algorithm variant that recomputes these values on-the-fly (\ALTO-OTF).
Moreover, we propose a heuristic to dynamically decide whether to reuse or recompute the intermediate results of tensor kernels based on the characteristics of the target data sets and tensor computations.
It is important to note that such pre-computation is different from prior memoization approaches~\cite{li2017model, kurt2022sparsity}, which use a non-trivial decision making process to reduce computations by \textit{reformulating} tensor operations and reusing intermediate results \textit{across modes}. In contrast, \ALTO-PRE uses easily calculable metrics to decide whether or not to reuse intermediate results \textit{within a mode}, and it performs the same tensor operations as \ALTO-OTF.

To demonstrate our adaptive memory management technique, 
Algorithm~\ref{fig:subproblem_algo} shows how the model update ($\mathbf{\Phi}$) kernel in CP-APR (Line~\ref{line:phi} from Algorithm~\ref{alg:cp-apr-mu}) is parallelized using the ALTO format.
The $\mathbf{\Pi}$ matrix from Line~\ref{line:pi} in Algorithm~\ref{alg:cp-apr-mu} calculates a dense matrix for a given mode $n$ that is the Khatri-Rao product (KRP) between all factor matrices \emph{excluding} the mode-$n$ factor matrix.
However, not every row of $\mathbf{\Pi}$ is required for a sparse tensor but only rows that correspond to nonzero elements are necessary and actually calculated, leading to a $\mathbf{\Pi}\in \mathbb{R}^{M\times R}$ matrix, where $M$ and $R$ are the number of nonzero elements and the decomposition rank, respectively.
In Algorithm~\ref{fig:subproblem_algo}, if we select to use  pre-computed $\mathbf{\Pi}$, the kernel reads in the $\mathbf{\Pi}$ matrix row corresponding to the nonzero element $x$ from memory (Line~\ref{line:pi-precompute});
otherwise, it computes the required KRP from the factor matrices (Line~\ref{line:otf-krp}) using the delineariezd coordinates. Pre-computing the $\mathbf{\Pi}$ matrix is simple;
in line~\ref{line:pi} of Algorithm~\ref{alg:cp-apr-mu} each nonzero element can calculate its Khatri-Rao product in parallel, using the equation from Line~\ref{line:otf-krp} in Algorithm~\ref{fig:subproblem_algo}.

Next, for each nonzero, the corresponding KRP row
is used to update the $\mathbf{\Phi}$ matrix, and the conflicting updates are resolved using our adaptive conflict resolution as detailed in $\S$\ref{sec:approach-cf}. Specifically, if there is high fiber reuse,
recursive tensor traversal is conducted and the update is made to the temporary scratchpad memory \MATRIX{Temp} (Line~\ref{line:sm}), which is later reduced (Lines~\ref{line:reduce-begin} to~\ref{line:reduce-end}) to decrease memory contention;
otherwise, output-oriented traversal is used to avoid synchronization and atomic operations are utilized to update the $\mathbf{\Phi}$ matrix  (Line~\ref{line:atomic}) only at the boundaries between different \ALTO partitions/threads.

\ALTO employs a simple heuristic for determining which algorithm variant (\ALTO-PRE or \ALTO-OTF) to use based on the fast memory size of hardware architectures as well as the fiber reuse and size of factor matrices of sparse tensors. Similar to our conflict resolution heuristic (illustrated in $\S$\ref{sec:approach-cf}), we use low fiber reuse to infer that on-the-fly computation of KRP is expensive, due to the cost of fetching data from memory. Hence, \ALTO decides to use pre-computation (\ALTO-PRE) when sparse tensors suffer from low fiber reuse and the size of their factor matrices is substantially larger than the fast memory size. Otherwise, the on-the-fly algorithm variant (\ALTO-OTF) is used because of its superior data locality and lower memory consumption.

%% file: text/results.tex
\section{Evaluation}\label{sec:perf}
We evaluate \ALTO-based tensor algorithms against the state-of-the-art sparse tensor libraries and representations in terms of parallel performance, tensor storage, and format generation cost. We conduct a thorough study of key tensor decomposition operations ($\S$\ref{sec:tensor_rank_decomposition}) and demonstrate the performance characteristics across the third and fourth generation of Intel Xeon Scalable processors, codenamed  Ice Lake~(ICX) and Sapphire Rapids~(SPR), respectively. 

\vspace*{-5pt}
\subsection{Experimental Setup}
\subsubsection{Platform}
The experiments were conducted on an Intel Xeon Platinum 8360Y CPU with \ac{ICX} micro-architecture, and an Intel Xeon Platinum 8470 CPU with \ac{SPR} micro-architecture. 
The \ac{ICX} system has 256 GiB main memory and it consists of two sockets, each with a 54\,\MiB ~L3 cache and 36 physical cores running at a fixed frequency of 2.4\,\GHZ ~for accurate measurements. 
The \ac{SPR} system also comprises two sockets with 52 physical cores each, and its frequency was fixed to 2.4\,\GHZ.
All cores within a socket share a 105\,\MiB ~L3 cache and the overall main memory in the \ac{SPR} server is 1\,\TB.
The experiments use all hardware threads ($72$ and $104$, respectively) on the target platforms. 
Both servers run AlmaLinux~8.8 Linux distribution and they are configured to enable ``Transparent Huge Pages'' and to support
two and four NUMA domains per socket on \ac{ICX} and \ac{SPR}, respectively.

The code is built using Intel C/C++ compiler (v2021.6.0) with the optimization flags \texttt{-O3} 
\texttt{-qopt-zmm-usage=high}
\texttt{-xHost} to fully utilize 
vector units. 
For performance counter measurements and thread pinning, we use the LIKWID tool suite~v5.3~\cite{likwid}.

\begin{table}[tb]
\scriptsize 
\centering
\caption{Characteristics of the target sparse tensor data sets. Non-negative count tensors are underlined.}
\vspace*{-2pt}
\label{tab:problems}
\begin{tabular}
{|p{0.92cm}|p{2.78cm}|p{0.78cm}|p{1.32cm}|p{0.8cm}|}
\hline
\textbf{Tensor} & \textbf{Dimensions} & \textbf{\#NNZs} & \textbf{Density} & \textbf{Fib. reuse}\\\toprule\hline
\textsc{\underline{lbnl}} & $1.6K\times4.2K\times1.6K\times4.2K\times868.1K$ & $1.7M$ & $4.2\times10^{-14}$ & Limited \\\hline 
\textsc{\underline{nips}} & $ 2.5K\times 2.9K\times 14K\times17$ & $3.1M$ & $1.8\times10^{-06}$& High\\\hline 
\textsc{\underline{uber}} & $ 183\times24\times 1.1K\times1.7K$&  $3.3M$ & $3.8\times10^{-04}$& High\\\hline 
\textsc{\underline{chicago}} & $ 6.2K\times24\times77\times32$& $5.3M$& $1.5\times10^{-02}$& High\\\hline 
\textsc{vast} & $165.4K\times11.4K\times2\times100\times89$& $ 26M$& $7.8\times10^{-07}$& High\\\hline 
\textsc{\underline{darpa}} & $22.5K\times22.5K\times23.8M$& $28.4M$& $2.4\times10^{-09}$& Limited \\\hline 
\textsc{\underline{enron}} & $ 6K\times5.7K\times244.3K\times1.2K$& $54.2M$& $5.5\times10^{-09}$& High \\\hline 
\textsc{\underline{lanl-2}} & $ 3.8K\times11.2K\times8.7K\times75.2K\times9$& $69.1M$& $1.9\times10^{-10}$& High \\\hline 
\textsc{nell-2} & $12.1K\times9.2K\times28.8K$& $76.9M$& $2.4\times10^{-05}$& High\\\hline 
\textsc{fb-m} & $23.3M\times23.3M\times166$ & $99.6M$& $1.1\times10^{-09}$& Limited\\\hline 
\textsc{flickr} & $319.7K\times28.2M\times1.6M\times731$& $112.9M$& $1.1\times10^{-14}$& Limited\\\hline 
\textsc{deli} & $532.9K\times17.3M\times2.5M\times1.4K$& $140.1M$& $4.3\times10^{-15}$& Medium\\\hline 
\textsc{nell-1} & $2.9M\times2.1M\times25.5M$& $143.6M$& $9.1\times10^{-13}$& Medium\\\hline 
\textsc{amazon} & $4.8M\times1.8M\times1.8M$& $1.7B$ & $1.1\times10^{-10}$& High\\\hline 
\textsc{patents} & $46\times239.2K\times239.2K$& $3.6B$ & $1.4\times10^{-03}$& High\\\hline 
\textsc{\underline{reddit}} & $8.2M\times177K\times8.1M$& $4.7B$& $4.0\times10^{-10}$& High\\\hline 
\end{tabular}
\end{table}

\subsubsection{Datasets}
The experiments consider a gamut of real-world tensor data sets with various characteristics. These tensors are often used in related works and they are publicly available in the FROSTT~\cite{frostt} and HaTen2~\cite{haten2_ICDE2015} repositories. Table~\ref{tab:problems} shows the detailed features of the target tensors, ordered by size, in terms of dimensions, number of nonzero elements (\#NNZs), and density. 
Additionally, the tensors are classified based on the average reuse of their fibers into high, medium, or limited reuse. We consider 
a given mode to have high reuse, if its fibers are reused more than eight times on average; when the fibers are reused between 
five to eight times, they have medium reuse; otherwise, the fibers suffer from limited reuse. Since TD operations access fibers along all modes, a tensor with at least one mode of limited/medium reuse is considered to have an overall limited/medium fiber reuse.
In the evaluation, we use all tensors and non-negative count tensors for CP-ALS and CP-APR experiments, respectively.

\begin{figure}[tb]
    \centering
    \begin{subfigure}[b]{0.49\linewidth} 
        \includegraphics{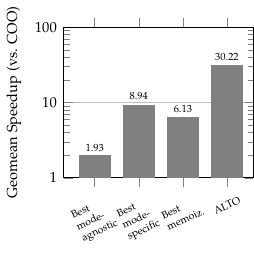} 
        \vspace*{-17pt}
        \caption{Performance on a 72-core ICX.}
        \label{fig:summary-ICX}
    \end{subfigure}
    \hfill
    \begin{subfigure}[b]{0.49\linewidth} 
        \includegraphics{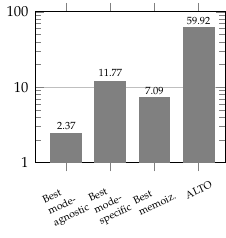} 
        \vspace*{-5pt}
        \caption{Performance on a 104-core SPR.}
        \label{fig:summary-SPR}
    \end{subfigure}
    \vspace*{-2pt}
    \caption{The performance of MTTKRP using \ALTO in comparison with an oracle selecting the best state-of-the-art variant for each implementation category for all tensors used in this paper.}
    \label{fig:summary}
\end{figure} 

\begin{figure*}[htb]
\centering

\begin{subfigure}[htp]{0.99\textwidth} 
        \includegraphics{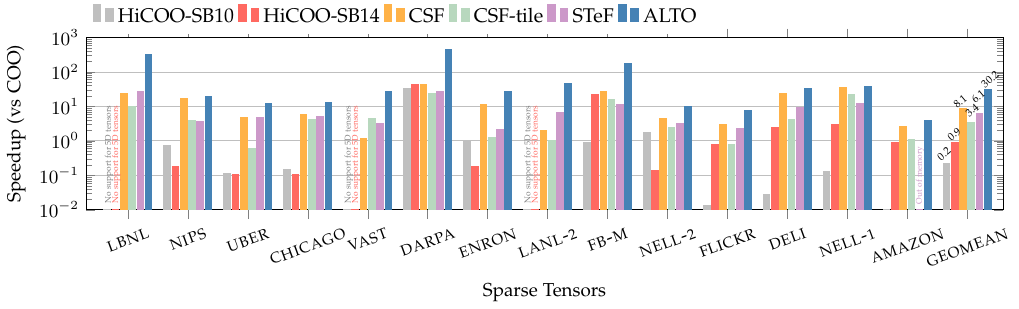}
	\vspace{-0.8em}
	\caption{Performance on a 72-core ICX.}
	\label{fig:CPDALS-ICX}
\end{subfigure}
\vspace*{2pt}

\begin{subfigure}[htp]{0.99\textwidth} 
        \includegraphics{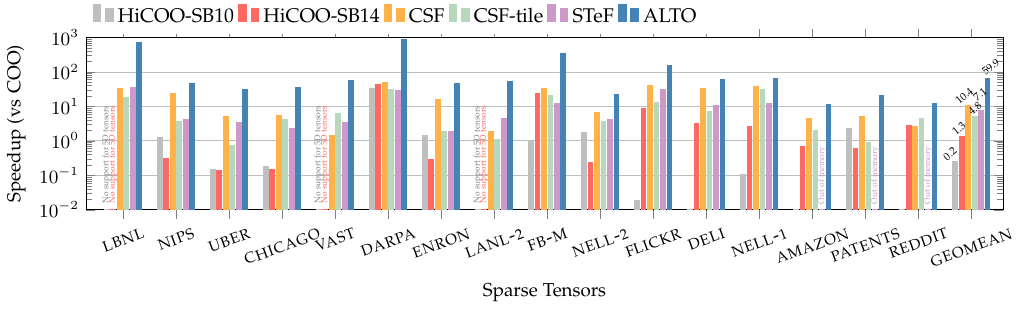}
	\vspace{-0.8em}
	\caption{Performance on a 104-core SPR.}
	\label{fig:CPDALS-SPR}
\end{subfigure}
\vspace*{-5pt}
	\caption{
    The overall parallel speedup of MTTKRP (all modes) using the different sparse tensor implementations compared to COO. 
 The sparse tensors are sorted in increasing order of their size (number of nonzero elements).
 }
	\label{fig:CPDALS}
 \vspace*{-5pt}
\end{figure*}

\vspace*{-2pt}
\subsubsection{Configurations}
We evaluate the proposed \ALTO format\footnote{Available at: \url{https://github.com/IntelLabs/ALTO}}, compared to the mode-agnostic COO and HiCOO formats~\cite{li2018hicoo, sparten} as well as the mode-specific CSF formats~\cite{Smith2015a, Smith2015, kurt2022sparsity}. Specifically, we use the latest code
of the state-of-the-art sparse tensor libraries 
for CPUs, namely, ParTI\footnote{Available at: \url{https://github.com/hpcgarage/ParTI}}, SPLATT\footnote{Available at: \url{https://github.com/ShadenSmith/splatt}}, and STeF\footnote{Available at: \url{https://github.com/HPCRL/STeF}} for normally distributed
data and SparTen\footnote{Available at: \url{https://github.com/sandialabs/sparten}} for non-negative count data.
On the ICX and SPR systems, we evaluate the target data sets that can fit in memory for all tensor libraries.  
We report the best-achieved performance across the different configurations of the COO format; that is, with or without thread privatization (which keeps local copies of the output factor matrix). For the HiCOO format, its performance and storage are highly sensitive to the block and superblock (SB) sizes, which benefit from tuning. Since the current HiCOO implementation does not auto-tune these parameters, 
we use a block size of 128 ($2^{7}$) and two superblock sizes of $2^{10}$ (``HiCOO-SB10'') and $2^{14}$ (``HiCOO-SB14'') according to prior work~\cite{sun2020sptfs}. 
We evaluate two variants of the mode-specific formats: CSF and CSF with tensor tilling (``CSF-tile''), both of which use $N$ representations
for an order-$N$ sparse tensor
to achieve the best performance.
For STeF, we use its data movement model to decide which results to memoize and reuse between modes based on the cache size. We report the best performance of STeF across the different cache configurations, which was achieved when setting the cache size to the size of L3 cache on the target ICX and SPR CPUs.

Similar to previous studies~\cite{Smith2015a, Smith2015, choi2018blocking}, the experiments use double-precision arithmetic and $64$-bit (native word) integers. To compute the CP-APR model for non-negative count data, we use $32$-bit integers to represent the input tensor values.
While the target data sets require a linearized index of size between $32$ and $80$ bits, we configured \ALTO to select the size of its linearized index to be multiples of the native word size (i.e., $64$ and $128$ bits) for simplicity. We use a decomposition rank $R=16$ for all experiments and set the maximum number of inner iterations ($l_{max}$) in CP-APR to $10$, as per prior work~\cite{Chi12Tensors}.

\begin{figure*}[htb]
	\centering
\vspace*{-10pt} 
\begin{subfigure}[htp]{0.9\textwidth}
\includegraphics{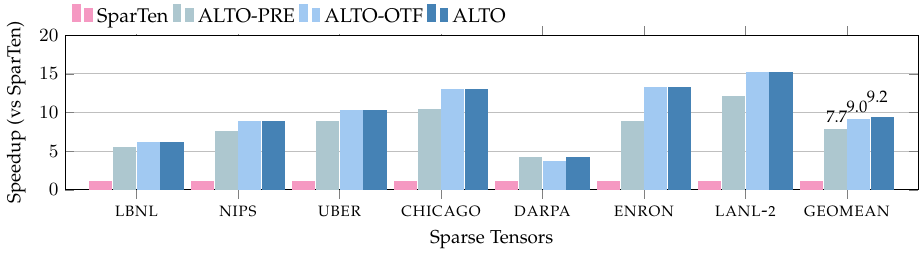}
	\vspace{-0.8em}
	\caption{Performance on a 72-core ICX.}
	\label{fig:CPAPR-ICX}
\end{subfigure}
\vspace*{-2pt} 

\begin{subfigure}[htp]{0.9\textwidth} 
\includegraphics{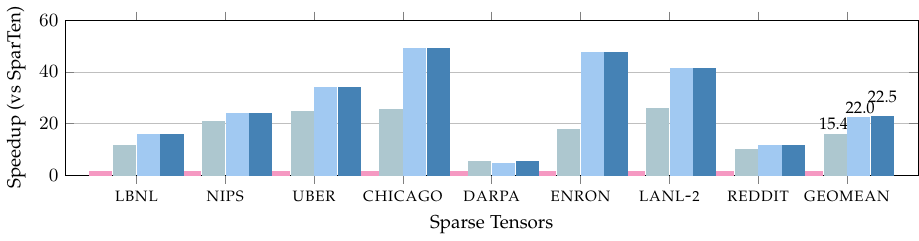}
	\vspace{-0.8em}
	\caption{Performance on a 104-core SPR.}
	\label{fig:CPAPR-SPR}
 \end{subfigure}
 \vspace*{-5pt}
	\caption{The achieved parallel speedup of the model update in CP-APR. The speedup is reported  compared to the state-of-the-art SparTen library. 
 \ALTO chooses between \ALTO-PRE and \ALTO-OTF using our adaptive memory management heuristic ($\S$\ref{subsec:phi}) to maximize performance.
 The sparse tensors are sorted in increasing order of their size (number of nonzero elements).}
	\label{fig:CPAPR}
\vspace*{-7pt}
\end{figure*}

\subsection{CP-ALS Performance}
We compare our \ALTO-based CP-ALS algorithm to a variety of CP-ALS implementations in the state-of-the-art \mbox{libraries} SPLATT, ParTI, and STeF for each tensor dataset. 
The implementations can be grouped into three categories: 1) mode-agnostic or general formats (COO, HiCOO-SB10, and HiCOO-SB14), which use one tensor copy, 2) mode-specific formats (CSF and CSF-tile), which keep multiple tensor copies (one per mode) for best performance, and 3) memoization techniques (STeF), which retain intermediate results across modes along with the tensor representation to reduce computations. 

Figures~\ref{fig:summary} and \ref{fig:CPDALS} show that \ALTO outperforms the best mode-agnostic and mode-specific formats as well as memoization schemes in terms of the speedup of tensor operations (MTTKRP on all modes).
In addition, the results indicate that \ALTO can effectively reduce synchronization and 
utilize the larger caches on SPR (relative to ICX) to further improve the performance compared to the state-of-the-art libraries. 
Specifically, \ALTO achieves $15.7\times$ and $25.3\times$ geometric mean (GEOMEAN) speedup on the ICX and SPR CPUs, respectively, compared to the best mode-agnostic formats. Although the mode-specific (CSF-based) 
formats require substantial storage to keep multiple tensor copies, \ALTO still delivers 
$3.4\times$ and $5.1\times$ geometric mean speedup on the ICX and SPR servers, respectively.
While memoization methods reduce computations, it comes at the cost of increasing memory traffic and consuming substantial amount of extra memory, which significantly limit their scalability. As a result, \ALTO realizes $4.9\times$ and $8.4\times$ geometric mean speedup over STeF on the ICX and SPR CPUs, respectively, while effectively handling all large-scale tensors that cause out-of-memory errors with STeF.
Furthermore, \ALTO shows 
scalable performance for the sparse tensors with high data reuse. 
Compared to its sequential version, \ALTO achieves up to $60\times$ and $80\times$ speedup on the 72-core ICX and 104-core SPR CPUs, respectively. 
For the other tensors (with limited/medium data reuse), \ALTO is bounded by memory bandwidth, and as a result it has an average speedup of around $20\times$ and $30\times$ on ICX and SPR, respectively.  
Conversely, the performance of the previous approaches is highly variable across data sets as it depends on the shape of sparse tensors as well as the spatial distribution of their nonzero elements rather than their inherent data reuse~\cite{alto_2021}. Specifically, the tree-based (CSF, CSF-tile, and STeF) and block-based (HiCOO-SB10/SB14) methods depend on grouping the nonzero elements into tensor slices and blocks for effective compression. As illustrated in Figure~\ref{fig:motivation}, finding \textit{balanced} clusters of nonzero elements in sparse tensors is highly unlikely. Thus, the prior  tree- and block-based techniques suffer from workload imbalance and inefficient compression, which in turn lead to limited parallel performance when scaling to a large number of cores.

\vspace*{-5pt}
\subsection{CP-APR Performance}
We compare our \ALTO-based CP-APR algorithm to the state-of-the-art SparTen library for non-negative count tensors.
SparTen uses a variant of the COO  format that keeps indexing as well as scheduling arrays for every tensor mode, which requires more than double the storage of COO~\cite{phipps2019software}. In addition, SparTen computes CP-APR using the traditional method that stores and then reuses intermediate results rather than recomputing them.
Hence, for large sparse tensors, such as \textsc{reddit}, SparTen fails to compute the CP-APR model on the ICX platform, even with 256 GiB of memory.
In contrast, \ALTO supports both recomputing (\ALTO-OTF) or storing and then reusing (\ALTO-PRE) intermediate results, which enables our CP-APR implementation to handle large-scale tensors. 

Figure~\ref{fig:CPAPR} shows the parallel performance of \ALTO-based CP-APR compared to SparTen on the ICX and SPR CPUs.
As the vast majority of time (more than 99\,\%) is spent in the model update ($\mathbf{\Phi}$) kernel (see Algorithm~\ref{fig:subproblem_algo}), the performance is evaluated based on the computation time of this tensor kernel.
Note that \ALTO-PRE and \ALTO-OTF in Figure~\ref{fig:CPAPR} represent the speedup achieved when the respective algorithms are used for all input tensors, and \ALTO represents the speedup achieved when our adaptive memory management heuristic ($\S$\ref{subsec:phi}) is used to choose between the two algorithms to maximize performance. 
Like CP-ALS, ALTO-based CP-APR realizes more scalable performance for tensors with high fiber reuse, and it  further improves the performance on SPR relative to ICX by reducing synchronization and leveraging the larger fast memories on SPR. 
Therefore, ALTO delivers substantial performance gains compared to the SparTen library, achieving $9.2\times$ and $22.5\times$ speedup on the ICX and SPR CPUs, respectively. 
Furthermore, as the tensor size and data reuse increase, our on-the-fly (\ALTO-OTF) algorithmic variant not only outperforms the traditional pre-computing approach (\ALTO-PRE) but also realizes better scalability, even when the intermediate results can fit in memory. 

\subsection{Performance Characterization}
Unlike prior compressed tensor 
formats, the parallel performance of \ALTO depends on the inherent data reuse of sparse tensors rather than the spatial distribution of their nonzero elements. 
To better understand the performance characteristics of \ALTO, we created a \Rlm~\cite{roofline:2009} for the SPR platform and collected performance counters across a set of representative parallel runs.
For the \Rlm, an upper performance limit \(\mathcal{P}\) is given by \(\mathcal{P} = \mathrm{min}(\mathcal{P}_{\mathrm{peak}}, B_{\mathrm{peak}} \times OI)\), where
\(\mathcal{P}_{\mathrm{peak}}\) is the peak performance, \(B_{\mathrm{peak}}\) is the peak memory bandwidth, and \(OI\) is the \emph{operational intensity} (i.e., the ratio of floating-point operations per byte).
Moreover, we enhance our \Rlm ~to consider the cache bandwidth. 
The L2/L3 cache and main memory bandwidth are measured using {\tt likwid-bench} from the Likwid tool suite.
Since L1 bandwidth measurements are error-prone, we use the theoretical L1 bandwidth of two cache lines per cycle per core. 
The peak performance, \(\mathcal{P}_{\mathrm{peak}}\), is calculated based on the ability of the cores to execute two fused multiply-add~(FMA) instructions on eight-element double precision vector registers~(due to the availability of AVX-512) per cycle.

\begin{figure}[htb]
    \centering
    \includegraphics{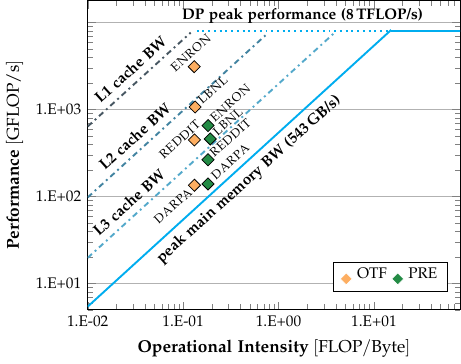}    
    \vspace*{-4pt}
    \caption{The parallel performance of the model update ($\mathbf{\Phi}$) in the CP-APR algorithm using \ALTO on a 104-core SPR system. Orange diamonds indicate on-the-fly computation (\ALTO-OTF), while green diamonds represent pre-computation (\ALTO-PRE).}
    \vspace*{-5pt}
    \label{fig:roofline}
\end{figure} 

\begin{figure*}[htb]
	\centering
        \includegraphics{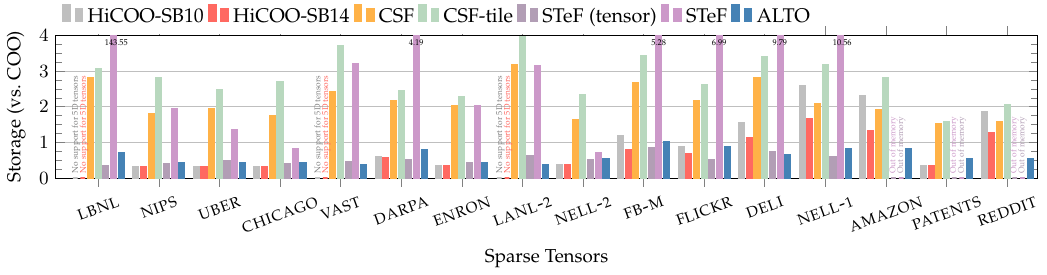}    
	\vspace{-0.9em}
	\caption{The tensor storage across the different formats relative to COO. The tensors are sorted in an increasing order of their size.}
	\label{fig:mem}
\end{figure*}

\begin{figure*}[htb]
	\centering
      \includegraphics{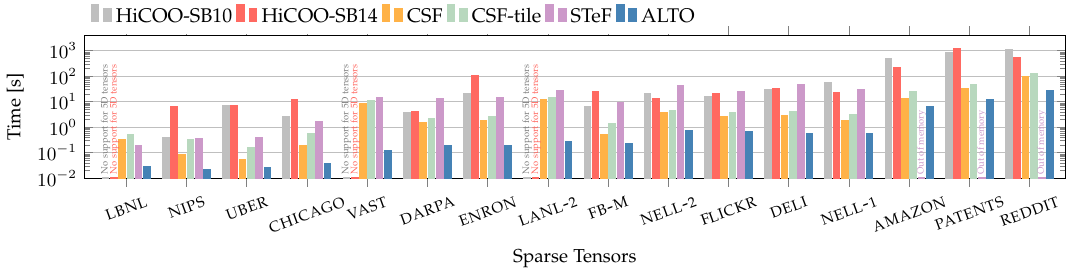}
	\vspace{-0.9em}
	\caption{The format construction cost on SPR in seconds. The sparse tensors are sorted in an increasing order of their size.}
    \vspace*{-5pt}
	\label{fig:perf-format}
\end{figure*}

Since prior work detailed the performance analysis of the CP-ALS algorithm and its MTTKRP kernel~\cite{alto_2021}, we focus on characterizing the performance of CP-APR in this study.  
Figure~\ref{fig:roofline} shows the performance of the parallel $\mathbf{\Phi}$-\ALTO kernel (Algorithm~\ref{fig:subproblem_algo}) for several representative tensors. 
To quantify the operational intensity, we calculated the required data movement from/to main memory as 
$l_{\textrm{avg}}mN(3R+RN+1)$ 
for on-the-fly computation (\ALTO-OTF) and $l_{\textrm{avg}}mN(3R+1)$ for pre-computation (\ALTO-PRE), where $l_{\textrm{avg}}$ is the average number of inner iterations, $m$ is the number of nonzero elements, $N$ is the number of modes, and $R$ is the decomposition rank.
We obtain the number of FLOPs required for the model update 
($\mathbf{\Phi}$)
by measuring hardware performance counters using {\tt likwid-perfctr} from the Likwid tool suite. 
The results indicate that although such memory-intensive computations suffer from low operational intensity, \ALTO can still exceed the peak main memory bandwidth by exploiting the inherent data reuse and by efficiently resolving update conflicts in caches.

Specifically, the Roofline plot shows the performance of \ALTO-OTF and \ALTO-PRE for two tensors with high data reuse (\textsc{enron} and \textsc{reddit}) and two tensor with limited data reuse (\textsc{lanl} and \textsc{darpa}).
For \textsc{enron},  we observe that \ALTO provides data access in a manner that allows the computation to be handled mainly from L1 and L2 cache. However, as a medium-sized tensor with high data reuse, it does not benefit from pre-computation and shows superior performance for the on-the-fly algorithm.
The \textsc{reddit} tensor, despite having good fiber reuse, is highly sparse and it is the largest tensor in our set of experiments (with 4.6\,billion nonzero elements).
This increases the memory pressure and effectively leads to more data accesses from slower memory,
which reduces the performance gap between \ALTO-OTF and \ALTO-PRE relative to the \textsc{enron} tensor.

While \textsc{lbnl} is extremely sparse and has limited data reuse, it is also the smallest of all tensors in the experiments.
This allows \ALTO to handle most of the data from the caches and to benefit from on-the-fly computation; however, the performance of \textsc{lbnl} is lower than denser tensors such as \textsc{enron}.
Finally, the \textsc{darpa} tensor, even though being similar in size to \textsc{enron}, has very limited fiber reuse (along mode-3). 
This leads to a significantly lower performance compared to any of the other tensors, yet the hybrid (recursive and output-oriented) tensor traversal of \ALTO
still captures some data reuse from caches and allows both \ALTO-PRE and \ALTO-OTF to realize superior performance, exceeding the main memory bandwidth.
For \textsc{darpa} we can observe a slightly better performance when using pre-computation.
Hence, the performance analysis indicates that \ALTO-PRE is especially relevant for large tensors that additionally show hyper-sparsity and limited data reuse. 

\vspace*{-4pt}
\subsection{Memory Storage}
Figure~\ref{fig:mem} details the relative storage of the different sparse tensor formats
compared to COO.
Due to its efficient linearization, as detailed in $\S$\ref{sec:approach-gen}, \ALTO requires less storage than the CSF, CSF-tile and raw (COO) formats for all investigated tensors.
The tree-based, mode-specific formats (CSF and CSF-tile) consume significantly more storage space than COO because they require multiple tensor representations for the different mode orientations.
While it can be beneficial for computation, imposing a tilling over the tensors (as done by CSF-tile) increases memory storage.
The memory consumption of the block-based formats (HiCOO-SB10/SB14) highly depends not only on the spatial distribution of the nonzero elements, but also on the block and superblock sizes. Compared to the COO format, HiCOO can reduce the memory footprint of tensors when the resulting blocks are relatively dense, i.e., the number of nonzero elements per block is high.
However, for hyper-sparse tensors such as \textsc{deli}, \textsc{nell-1}, \textsc{amazon}, and \textsc{reddit}, HiCOO requires more storage by up to a factor of $2.6$. 
While the tensor format of STeF~(``STeF (tensor)'' in Figure~\ref{fig:mem}) only requires storage on-par or even smaller than ALTO, the additional memory needed for memoization leads to a higher memory footprint~(``STeF'') compared to ALTO in all cases.

\subsection{Format Generation Cost}\label{sec:perf-format} 
Figure~\ref{fig:perf-format} details the generation cost of the different sparse tensor representations from a sparse tensor in the COO format on the SPR platform. 
Instead of processing nonzero elements in a multi-dimensional form as HiCOO and CSF, \ALTO works on a linearized representation that needs substantially lower number of comparison operations to sort nonzero elements.
Furthermore, the HiCOO formats require additional clustering of elements based on their multi-dimensional coordinates, as well as scheduling of the blocks and superblocks for avoiding conflicts, while STeF requires additional sorting of the nonzero elements along a specific mode order for best performance.
Thus, \ALTO achieves substantial geometric mean speedup for format generation compared to HiCOO-SB10~($50\times$), HiCOO-SB14~($75\times$), CSF-tile~($10\times$), CSF~($6\times$), and STeF~($44\times$).

%% file: text/related.tex
\section{Related Work} \label{sec:related}
Our mode-agnostic \ALTO format was motivated by the linearized coordinate (LCO) format~\cite{harrison2018high}, which also flattens sparse tensors but in a mode-specific way, i.e., along a given mode orientation. 
Hence, LCO requires either multiple tensor copies or permuting tensors for efficient computation. 
Additionally, the authors limit their focus to sequential algorithms, and it is not clear how LCO can be used to efficiently execute parallel sparse tensor computations.

Researchers proposed variants~\cite{Liu2017, phipps2019software} of the COO format to reduce the synchronization overhead using \textit{mode-specific} scheduling arrays. The state-of-the-art SparTen library~\cite{sparten} uses a COO variant~\cite{phipps2019software} to accelerate the decomposition of non-negative count tensors across different hardware architectures. However, these COO variants adversely affect the input data locality and lead to random access of the nonzero elements~\cite{phipps2019software}, especially for sparse tensors with high data reuse. 
Moreover, keeping fine-grained scheduling information for \textit{all tensor modes} can more than double the memory consumption, compared to the COO format~\cite{phipps2019software}. 

The popular SPLATT library~\cite{Smith2015} leverages the CSF format~\cite{Smith2015, Smith2015a} to decompose sparse tensors on multi-core CPUs. However, this mode-specific compressed format requires multiple tensor copies for best performance. In addition, CSF packs the nonzero elements into coarse-grained tensor slices and fibers, which limits its scalability on massively parallel architectures. 
To improve the performance on GPUs, recent CSF-based formats~\cite{nisa2019load, nisa2019efficient} expose more balanced and fine-grained parallelism but at the expense of substantial synchronization overheads and expensive preprocessing and format generation costs.

Alternatively, the ParTI library uses the mode-agnostic, block-based HiCOO format~\cite{li2018hicoo} to decompose sparse tensors using only one tensor copy. Yet, HiCOO is highly sensitive to the characteristics of sparse tensors as well as the block size. Due to the irregular (skewed) data distributions in sparse tensors, the number of nonzero elements per block varies widely across HiCOO blocks, even after expensive mode-specific tensor permutations which in practice further increase workload imbalance~\cite{li2019efficient}. 
As a result, when tensors are highly sparse, HiCOO can consume more storage than the COO format~\cite{li2018hicoo}. 
Moreover, using small data types for indexing nonzero elements within a block can end up under-utilizing the compute units and memory bandwidth in modern parallel architectures~\cite{Abel19a, mittal2017survey}, which are optimized for wide memory transactions~\cite{ghose2019demystifying}.
 
STeF~\cite{kurt2022sparsity} leverages the mode-specific CSF format to accelerate all-modes MTTKRP by memoization of partial MTTKRP results. Nevertheless, the additional space required for memoization can be more than double the memory storage of the sparse tensor and factor matrices, which limits STeF applicability to small- and medium-scale tensors~\cite{kurt2022sparsity}.
SpTFS~\cite{sun2020sptfs} utilizes machine learning~\cite{zhao2018bridging, xie2019ia} to predict the best of COO, HiCOO, and CSF formats to compute MTTKRP for a given sparse tensor. FLYCOO~\cite{wijeratne2023accelerating, wijeratne2023dynasor} extends the COO format to memory-constrained platforms (such as FPGAs) by processing a tensor into small equal-sized shards. However, FLYCOO requires dynamic mode-specific tensor remapping/reordering and it needs more storage than COO to keep sharding information for every mode.

%% file: text/conclusion.tex
\section{Conclusion} \label{sec:conclusion}
To overcome the limitations of existing sparse tensor formats, this work introduced \ALTO, a compact mode-agnostic format to efficiently encode higher-order tensors of irregular shapes and data distributions.
Thanks to their adaptive tensor traversal and superior workload balance and data reuse,
our \ALTO-based parallel algorithms for decomposing normally distributed data (CP-ALS) and non-negative count data (CP-APR) delivered an order-of-magnitude speedup over the best mode-agnostic formats while requiring $\sim 50\%$ of COO storage. 
Moreover, \ALTO achieved $5.1\times$ and $8.4\times$ geometric mean speedup over the best mode-specific and memoization methods, respectively, while needing between $14\%$ to $25\%$ of their overall storage.
Our future work will investigate distributed-memory platforms and other common sparse tensor algorithms, besides tensor decomposition.

\balance